\documentclass[10pt, journal]{IEEEtran}

\usepackage[utf8]{inputenc}%(only for the pdftex engine)

\usepackage[utf8]{inputenc}
\usepackage{float} % for image placement
\usepackage{amsmath}
\usepackage{amssymb}
\usepackage{stmaryrd}
\usepackage{multirow}
\usepackage[utf8]{inputenc}
\usepackage{color}
\usepackage[linesnumbered,ruled,noline,procnumbered]{algorithm2e}
\usepackage{enumitem}

\SetAlgoProcName{Protocol}{List of protocols}

\usepackage{tcolorbox}

\tcbuselibrary{skins,breakable}
\newtcolorbox{functionality}[2][]{%
  enhanced,
  title        = {#2},
  attach boxed title to top left={xshift=+3mm,yshift*=-3mm},
  breakable    = true,
  colback      = blue!5,
  colframe     = blue!35!black,
  fonttitle    = \bfseries,
  colbacktitle = blue!15!white,
  coltitle     = black,
  #1
}

\newtcolorbox{mynote}[1][]{%
  enhanced jigsaw, % better frame drawing
  breakable        = true,
  borderline west  ={2pt}{0pt}{red}, % straight vertical line at the left edge
  sharp corners, % No rounded corners
  boxrule          =0pt, % no real frame,
  fonttitle        ={\large\bfseries},
  coltitle         ={black},  % Black colour for title
  colback          =yellow!25,
  title            ={Review:\ },  % Fixed title
  attach title to upper, % Move the title into the box
  #1
}
\newtcbox{\xmybox}[1][red]{on line,
arc=7pt,colback=#1!10!white,colframe=#1!50!black,
before upper={\rule[-3pt]{0pt}{10pt}},boxrule=1pt,
boxsep=0pt,left=6pt,right=6pt,top=2pt,bottom=2pt}

\newcommand{\lss}{\ensuremath{ [\![ }}
\newcommand{\rss}{\ensuremath{ ]\!] }}

\newcommand{\ptrunc}{\ensuremath{\pi_{\mathsf{trunc}}}\xspace}
\newcommand{\pdecomp}{\ensuremath{\pi_{\mathsf{decomp}}}\xspace}
\newcommand{\pdecompopt}{\ensuremath{\pi_{\mathsf{decompOPT}}}\xspace}
\newcommand{\ptoq}{\ensuremath{\pi_{\mathsf{2toQ}}}\xspace}

\newcommand{\pbtx}{\ensuremath{\pi_{\mathsf{BTX}}}\xspace}
\newcommand{\pgeq}{\ensuremath{\pi_{\mathsf{GEQ}}}\xspace}
\newcommand{\peq}{\ensuremath{\pi_{\mathsf{EQ}}}\xspace}
\newcommand{\pminmax}{\ensuremath{\pi_{\mathsf{minmax}}}\xspace}

\newcommand{\pmul}{\ensuremath{\pi_{\mathsf{DM}}}\xspace}
\newcommand{\pmmul}{\ensuremath{\pi_{\mathsf{DMM}}}\xspace}

\newcommand{\pdisc}{\ensuremath{\pi_{\mathsf{DISC}}}\xspace}

\newcommand{\pRF}{\ensuremath{\pi_{\mathsf{RF}}}\xspace}
\newcommand{\pXT}{\ensuremath{\pi_{\mathsf{XT}}}\xspace}

\newcommand{\pSIDT}{\ensuremath{\pi_{\mathsf{SID3T}}}\xspace}
\newcommand{\fti}[1]{\ensuremath{\mathcal{F}^{\mathcal{D}_{#1}}_{\mathsf{TI}}}\xspace}
\newcommand{\fmmul}{\ensuremath{\mathcal{F}_{\mathsf{DMM}}}\xspace}
\newcommand{\fconv}{\ensuremath{\mathcal{F}_{\mathsf{2toQ}}}\xspace}
\newcommand{\fbtx}{\ensuremath{\mathcal{F}_{\mathsf{BTX}}}\xspace}
\newcommand{\fgeq}{\ensuremath{\mathcal{F}_{\mathsf{GEQ}}}\xspace}
\newcommand{\feq}{\ensuremath{\mathcal{F}_{\mathsf{EQ}}}\xspace}
\newcommand{\fXT}{\ensuremath{\mathcal{F}_{\mathsf{XT}}}\xspace}
\newcommand{\fSIDT}{\ensuremath{\mathcal{F}_{\mathsf{SID3T}}}\xspace}
\newcommand{\fminmax}{\ensuremath{\mathcal{F}_{\mathsf{minmax}}}\xspace}
\newcommand{\fdisc}{\ensuremath{\mathcal{F}_{\mathsf{DISC}}}\xspace}
\newcommand{\fdiscdt}{\ensuremath{\mathcal{F}_{\mathsf{DISC+DT}}}\xspace}
\newcommand{\fdiscrf}{\ensuremath{\mathcal{F}_{\mathsf{DISC+RF}}}\xspace}
\newcommand{\Zqm}[2]{\ensuremath{\mathbb{Z}_q^{#1\times #2}}\xspace} 
\newcommand{\shareq}[1]{\ensuremath{\llbracket{#1}\rrbracket_{_q}}\xspace}
\newcommand{\sharetwo}[1]{\ensuremath{\llbracket{#1}\rrbracket_{_2}}\xspace}

\newcommand{\s}{\ensuremath{\mathcal{S}}\xspace}
\newcommand{\env}{\ensuremath{\mathcal{Z}}\xspace}
\newcommand{\adv}{\ensuremath{\mathcal{A}}\xspace}
\newcommand{\F}{\ensuremath{\mathcal{F}}\xspace}

\def\getsr{\stackrel{{\scriptscriptstyle\$}}{\leftarrow}}

\begin{document}

\title{Privacy-Preserving Training of Tree Ensembles over Continuous Data}

\author{Samuel Adams, Chaitali Choudhary, Martine De Cock, Rafael Dowsley, David Melanson, Anderson~C.~A.~Nascimento, Davis Railsback, Jianwei Shen
 \thanks{Samuel Adams, Chaitali Choudhary, Martine De Cock, David Melanson, Anderson~C.~A.~Nascimento and Davis Railsback are with the School of Engineering and Technology, University of Washington Tacoma, Tacoma, WA 98402. Emails: \{sdadams,cc201,mdecock,mence40,andclay,drail\}@uw.edu}
\thanks{Martine De Cock is a Guest Professor at Dept.~of Applied Mathematics, Computer Science, and Statistics, Ghent University}
\thanks{Rafael Dowsley is with the Faculty of Information Technology, Monash University, Clayton, Australia. Email: rafael.dowsley@monash.edu}
\thanks{Jianwei Shen is with the Department of Computer Science, University of Arizona. Email: sjwjames@email.arizona.edu}
}

\maketitle

\begin{abstract}
Most existing Secure Multi-Party Computation (MPC) protocols for privacy-preserving training of decision trees over distributed data assume that the features are categorical. In real-life applications, features are often numerical. The standard ``in the clear'' algorithm to grow decision trees on data with continuous values requires sorting of training examples for each feature in the quest for an optimal cut-point in the range of feature values in each node. Sorting is an expensive operation in MPC, hence finding secure protocols that avoid such an expensive step is a relevant problem in privacy-preserving machine learning. 
In this paper we propose three more efficient alternatives for secure training of decision tree based models on data with continuous features, namely: (1) secure discretization of the data, followed by secure training of a decision tree over the discretized data; (2) secure discretization of the data, followed by secure training of a random forest over the discretized data; and (3) secure training of extremely randomized trees (``extra-trees'') on the original data. Approaches (2) and (3) both involve randomizing feature choices. In addition, in approach (3) cut-points are chosen randomly as well, thereby alleviating the need to sort or to discretize the data up front. We implemented all proposed solutions in the semi-honest  setting with additive secret sharing based MPC.  
In addition to mathematically proving that all proposed approaches are correct and secure, we experimentally evaluated and compared them in terms of classification accuracy and runtime. We privately train tree ensembles over data sets with 1000s of instances or features in a few minutes, with accuracies that are at par with those obtained in the clear. This makes our solution orders of magnitude more efficient than the existing approaches, which are based on oblivious sorting.
\end{abstract}


\IEEEpeerreviewmaketitle

%%%%%%%%%%%%%%%%%%%%%%%%%%%%%%%%%%%%%%%%%%%%%%%%%%%%%%%%%%%%%%%%%
%
%   1. INTRODUCTION
%
%%%%%%%%%%%%%%%%%%%%%%%%%%%%%%%%%%%%%%%%%%%%%%%%%%%%%%%%%%%%%%%%%

\section{Introduction}
%Privacy-Preserving Machine Learning (PPML) uses a combination of machine learning (ML) and cryptographic techniques. ML
%, one of the most popular research fields in computer science, 
%has a wide variety of applications, many of which
%Many of these applications 
%use individuals’ personal data or data originating from different sources with different security policies. E.g., multiple hospitals may want to train a model for a particular disease diagnosis based on their combined data; however, at the same time these hospitals may need to maintain
%the privacy of their patients. 
Secure Multi-Party Computation (MPC) is a powerful tool for achieving Privacy-Preserving Machine Learning (PPML). For example, instantiating PPML based on Secure Multi-Party Computation \cite{CDN2015} enables multiple parties to work together to train an ML model over their combined data, without any of the parties learning anything about each other's data.

Recent advances in MPC-based protocols for \textit{training} of ML models over distributed data are primarily focused on secure 
 training of neural network architectures  \cite{mohassel2017secureml,wagh2019securenn,idash,guo2020secure}. While deep learning is state-of-the-art for tasks that relate to perception, such as computer vision and natural language processing, in domains with structured information, the best results are often obtained with tree ensemble methods, such as random forests and boosted decision trees \cite{dietterich2000ensemble}. The latter also have the advantages of being faster to train and being easier to interpret. Advances on MPC-based training of tree based classifiers are fairly limited. While there is work on \textit{secure inference} with pre-trained tree ensembles \cite{fritchman2018privacy}, work on \textit{secure training} itself is limited to the training of individual decision trees (DTs). Several authors have proposed a secure version of Quinlan's ID3 algorithm \cite{quinlan1986induction} for training DTs with \textit{categorical} features (features) \cite{lindell2000privacy,vaidya2005privacy,samet2008privacy,de2014practical}.
Existing proposals for training DTs with \textit{continuous} features \cite{xiao2006privacy,shen2009privacy,behera2011privacy,escudero2020}
are based on Quinlan's C4.5 algorithm \cite{quinlan2014c4}, an algorithm that involves sorting, a time-consuming operation in the {MPC} setting.

Secure DT learning with {MPC} is challenging for a variety of reasons. For algorithms in general to be secure in MPC, measures must be taken to ensure that the number of executions of instructions is not dependent on specific values of the input, because that in itself could leak information. In the context of ML algorithms, where models are trained privately and not revealed to the parties, this means for instance that one should not rely on early stopping conditions, or on control flow logic. Furthermore, for efficiency considerations, dependency on previous multiplication results should be minimized, and operations like division, analytic function evaluation, and integer logic should be avoided where possible.
All of these Achilles' heels of MPC are inherent requirements of traditional DT training algorithms. Indeed, trees are grown recursively, which implies several layers of dependency on previous results. Furthermore, the ``decision'' component of a tree indicates a stopping condition has been met (thereby potentially revealing information, if one is not careful), and the information gain metric for greedy selection of splitting features requires computing division and analytic functions. In this paper, we introduce novel methods to contend with these requirements.  

We propose three alternative strategies for secure training of tree based models over data with continuous feature values, none of which requires sorting of feature values. Two of our approaches rely on privacy-preserving discretization of the range of feature values. After this step, any of the existing algorithms for secure training of a DT over categorical data can be used. In our case, we use the SID3T training algorithm proposed by de Hoogh et al.~\cite{de2014practical}. This constitutes our first approach.

Next, we present a novel protocol for secure training of a random forest (RF) over data with categorical values, by extending de Hoogh et al.'s secure DT training algorithm with a protocol for privacy-preserving random feature selection. Combined with the secure discretization protocol from approach 1, this allows us to securely train RFs over data with continuous values, constituting our second approach.

When growing a tree for data with continuous features (e.g.~\textit{price}), the standard ``in the clear'' C4.5 algorithm sorts the feature values to look for a cut-off point (e.g.~$\leq 100\$$) that will reduce class label impurity the most at the next level of the tree. To circumvent expensive secure sorting operations, as our third approach, we propose secure training of tree ensembles based on randomized choices for the cut-off points. In the clear, this idea, combined with randomized feature selection, is known as \textit{extremely randomized trees}. Such ``extra-trees classifiers'' achieve state-of-the-art accuracy and are fast to train over data sets with numerical features \cite{geurts2006extremely}.
%
%The third approach, relies on the random selection of thresholds used for comparisons in the decision tree nodes. 
%
Summarizing, in this paper: 

\begin{itemize}[leftmargin=*,topsep=0pt]
    \item We propose an {MPC}-based protocol \pdisc for privacy-preserving discretization of a range of continuous feature values, in scenarios where the feature values are distributed across different parties.
    \item We present the first {MPC}-based protocol \pRF for training of a random forest (RF). 
    %over distributed data with categorical values. \olive{Should this "continuous" instead of "categorical"? We take continuous data, discretize it, effectively turning it into categorical values, and then train SID3T on it, but the original data was still continuous, and this discretization process would likely not do well on data that was originally categorical (suggests ordering) - David}
    \item We propose the first {MPC}-based protocol \pXT for training of an extra-trees classifier (XT).
    %over distributed data with continuous values.
    \item As a side result, we propose several improvements and optimizations to important building blocks of privacy-preserving machine learning protocols such as secure comparisons.
\end{itemize}
Combined, these protocols allow to train tree based models over distributed data with continuous values in a variety of ways: (1) secure discretization of the data with \pdisc, followed by secure training of a DT over the discretized data with the \pSIDT protocol from de Hoogh et al.~\cite{de2014practical}; (2) secure discretization of the data with \pdisc, followed by secure training of a random forest over the discretized data with \pRF; and (3) secure training of extremely randomized trees on the original data with \pXT.
All our approaches accommodate scenarios in which the trained ML models have to remain private, i.e.~secret shared across the parties, as well as scenarios in which the trained ML models are disclosed at the end. Furthermore, all our approaches work in scenarios where the data is horizontally partitioned (each party has some of the rows or instances), scenarios where the data is vertically partitioned (each party has some of the columns or features), and even in scenarios where each computing party only has secret shares of the data to begin with.

We compare the accuracy and the runtime of the proposed approaches on publicly available benchmark data sets with 1000s of instances or features. Our solutions are simple, and, for the most part, use building blocks that already exist in the literature. Importantly, our solutions work. We obtain accuracies that are at par with those that can be obtained with existing sorting based MPC protocols for training of tree based models, while being much faster. Our results show that a full sort is not necessary for MPC-based training of tree based models.\\ 
%based on an implementation of the protocols in the Lynx framework.\footnote{https://bitbucket.org/uwtppml}  

%While most of our building blocks already exist in the literature, the novelty of our work comes from asking the question of how one can efficiently implement and apply these building blocks to the problem of training continuous data based on decision trees and tree ensembles. 

\noindent
\textbf{Related Work.}
The majority of the existing work on the training of DTs is for categorical data only \cite{lindell2000privacy,vaidya2005privacy,samet2008privacy,de2014practical}. 
In a preliminary attempt to extend MPC-based privacy-preserving DT training algorithms to continuous variables, Xiao et al.~\cite{xiao2006privacy} proposed a straightforward adaptation of the C4.5 algorithm
%to the privacy-preserving setting. They proposed to use a 
with a privacy-preserving bubble sort algorithm, which is not practical. Moreover, the authors did not give security proofs for their proposed protocols. Shen et al.~\cite{shen2009privacy} extended the result presented in \cite{xiao2006privacy} to vertically partitioned data. 

More recently, Abspoel et al.~\cite{escudero2020} proposed a new MPC-based adaptation of the C4.5 algorithm, using sorting networks to obliviously presort the feature values. Subsequently they consider each feature value as a candidate cut-off point, and compute the Gini index for each. They estimate that training one DT of depth 4 on a data set with 8192 instances and 11 features would take slightly over 8 min when run on three m5d.2xlarge EC2 instances connected via a LAN. They extrapolate from this that training an ensemble of 200 such trees, each over a sample of 8192 instances and 11 features drawn from a much larger data set of training instances and features, could be done in less than 28 hours. Their solution does not include a mechanism for performing bagging (selection of the instances used in each tree) and subspace sampling (selection of the features used in each tree) in a privacy-preserving manner. In this paper, we propose MPC protocols to this end in what is, to the best of our knowledge, the first end-to-end protocol for MPC-based training of random forests. We also present a working implementation that allows to train accurate tree ensembles over data sets with 1000s of instances or features in a matter of minutes, i.e.~our solution is much faster than the existing ones. These improvements in efficiency stem from the fact that we do not sort feature values, and from the fact that we need to perform far less Gini index computations, because we perform a much more straightforward discretization of the data, while maintaining high accuracy.

Our work on MPC protocols for privacy-preserving training of random forests (RFs) and extra-trees classifiers (XTs) was carried out independently from simultaneous work on MPC-based training of gradient boosted decision tree models (XGBoost) \cite{cryptoeprint:2021:432}. RFs and XTs are inherently different from XGBoost. Whereas for XGBoost an ensemble of trees is trained in sequence by adding, at each step, the tree with the greatest accuracy improvement, for RFs and XTs, many trees are trained independently on different random subsamples of the data. RFs,  XTs, and XGBoost are all popular tree ensemble methods in data science that may outperform one another in predictive accuracy depending on the data set and the task at hand.

%To the best  of our knowledge, our protocols are the first ones that attack the problem of training DTs and DT ensembles on continuous data while avoiding a full sort of the data set.   

%%%%%%%%%%%%%%%%%%%%%%%%%%%%%%%%%%%%%%%%%%%%%%%%%%%%%%%
%
%      2. PRELIMINARIES
%
%%%%%%%%%%%%%%%%%%%%%%%%%%%%%%%%%%%%%%%%%%%%%%%%%%%%%%%
\section{Preliminaries}\label{SEC:PRELIM}
\noindent
\textbf{Security setting.}
We consider \textit{honest-but-curious, static adversaries}, as is common in {MPC}-based PPML (see e.g.~\cite{de2014practical,IEEETDSC:CDHK+17}). 
An honest-but-curious adversary (also known as passive or semi-honest adversary) follows the instructions of the protocol, but tries to gather additional information. Secure protocols prevent the latter. A static adversary chooses the parties that he wants to corrupt before the protocol execution. Our security model and proofs are in Appendix \ref{sec:security}.\\

\noindent
\textbf{Fixed point representation.}
The protocols proposed in this paper are designed for training of DT based models on data with continuous feature values. The training data consists of a set $S$ of training examples $\langle(x_1,x_2,\ldots,x_f),y\rangle$.
The feature values $x_1, x_2,\ldots,x_f$ are real numbers, while the class label $y$ is categorical. The goal is to learn a function from the data that maps previously unseen feature values to a corresponding class label, e.g.~to determine whether a patient has a disease or not based on blood pressure, temperature etc. The kind of functions that we consider in this paper are DT based models. Our assumption is that, instead of residing in one place, the data set $S$ is distributed across multiple data owners. 

%Before the start of a secure training protocol, two parties, nicknamed \textit{Alice} and \textit{Bob} each have part of the data set $S$ on which the classifier should be trained. The data can be horizontally partitioned, meaning that Alice and Bob each have their own rows (instances); it can also be vertically partitioned, meaning that Alice and Bob each own some of the columns (features).
During the execution of the protocols however, operations are performed on additive shares in a ring $\mathbb{Z}_q$, for some appropriately chosen integer $q$. In this work we mostly   use $q=2^\lambda$. The feature values $x$ in $\mathbb{R}$ first need to be converted into values $Q(x)$ in $\mathbb{Z}_{2^\lambda}$ by the data owners. To this end we use a fixed point representation with two's complement for negative numbers: 
\begin{equation}
\label{eq:fixedpoint}
Q(x) = 
\begin{cases} 
      2^\lambda - \left \lfloor{ 2^a \cdot |x|  }\right \rfloor & \mbox{if\ } x < 0 \\
      \left \lfloor{ 2^a \cdot x  }\right \rfloor & \mbox{if\ } x \geq 0
\end{cases}
\end{equation}

When converting $Q(x)$ into its bit representation, it consists of $\lambda$ bits in total. The first $a$ bits from the right hold the fractional part of $x$, the next $b$ bits represent the non-negative integer part of $x$, and the most significant bit (MSB) represents the sign (positive or negative).  It is important to choose  $\lambda$ large enough to be able to represent the largest numbers produced during the protocols. When multiplying fixed point numbers, the number of fractional bits doubles and must be truncated to remain in the proper range. Therefore, $\lambda$ should be chosen to be at least $2(a+b)$. 
It is also important to choose $b$ large enough to represent the maximum possible value of the integer part of all $x$'s. 

After the conversion, each data owner secret shares the feature values on its end. In general, a number $z$ in $\mathbb{Z}_q$ is split in $m$ shares by picking $z_1, z_2, \ldots, z_m \in \mathbb{Z}_q$ uniformly at random subject to the constraint that $z = \sum_{i}z_i \mod{q}$. 
All computations are modulo $q$ and the modular notation is henceforth omitted for conciseness.
We denote this secret sharing by $[\![z]\!]_q$, which can be thought of as a shorthand for $(z_1,z_2,\ldots,z_m)$. For the case of $q=2^\lambda$, we simplify the notation to $[\![z]\!]$. Note that no information about the secret value $z$ is revealed by any proper subset of the $m$ shares, but the secret shared value can be trivially revealed by combining all shares. The class label of training examples is secret shared in the same manner as the feature values.\\ 

\begin{figure*}
    \centering
    \includegraphics[width=0.65\textwidth]{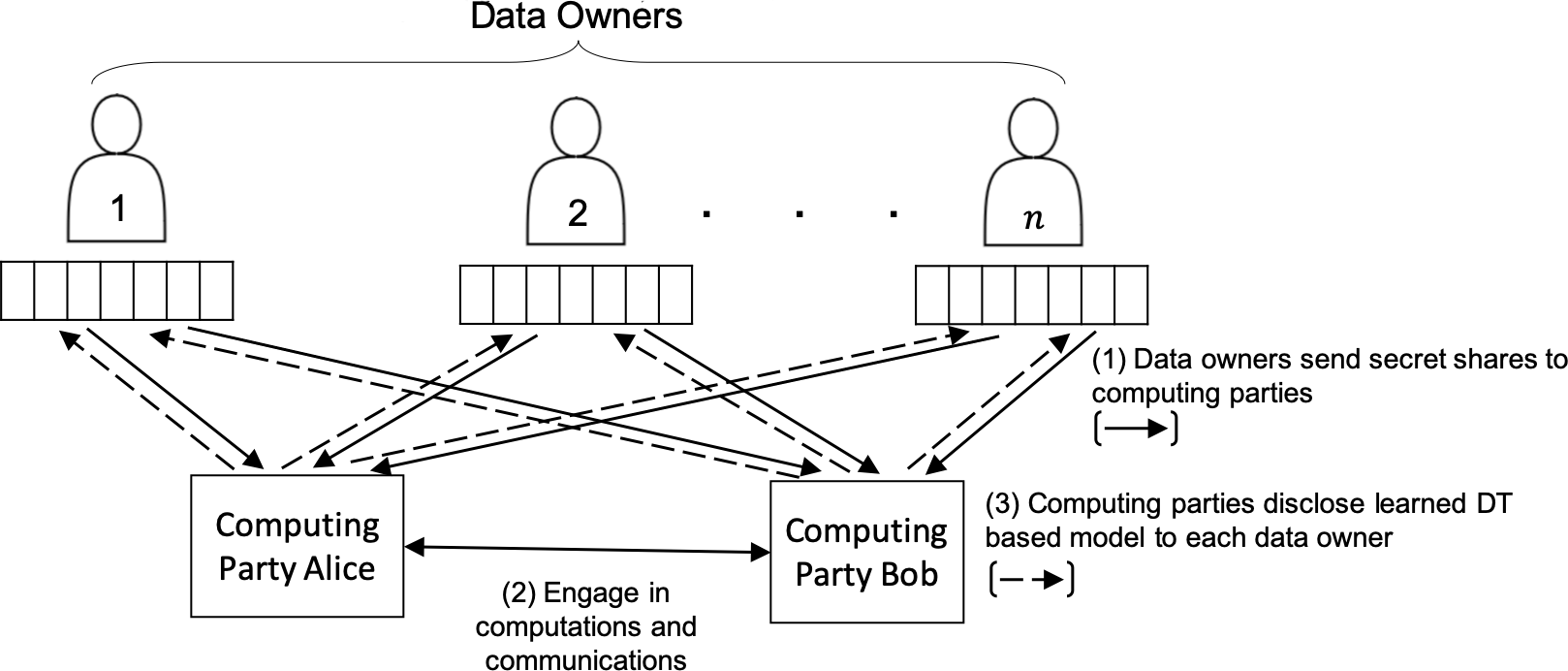}
    \caption{Overview of MPC based secure decision tree (DT) based model training. 
    Each of $n$ data owners secret shares their own training data between two computing parties. The computing parties engage in computations and communications to train a DT based model, which is at the end revealed to the data owners. As an alternative to Step (3), our protocols also allow for the learned model to remain hidden, i.e.~secret shared between the computing parties. In this case, when a new instance has to be classified, the new instance is secret shared across the computing parties also, who then work together to each obtain secret shares of the class label.}
    \label{fig:diagram}
\end{figure*}

\noindent
\textbf{Operations on secret shared values.}
For ease of explanation we will from now on focus on the case of $m=2$ and refer to the computing parties that receive the respective shares as \textit{Alice} and \textit{Bob}. Figure~\ref{fig:diagram} illustrates this scenario where $n$ data owners secret share their data with $m=2$ computing parties. It should be noted that the techniques proposed in this paper can be extended to work for $m > 2$ computing parties with adequate substitutions for the 2PC-exclusive subprotocols \ptrunc (see below), \peq and \pgeq (see Section \ref{SEC:COMP}). 

Given secret shared values $[\![x]\!]_q$ and $[\![y]\!]_q$, and a constant $c$, Alice and Bob can trivially perform the following operations locally:
\begin{itemize}[leftmargin=*,topsep=0pt]
  \item Addition ($z=x+y$): Alice and Bob just add their local shares of $x$ and $y$. This operation will be denoted by 
  $[\![z]\!]_q \leftarrow[\![x]\!]_q+[\![y]\!]_q$.
  \item Subtraction ($z=x-y$): Alice and Bob subtract their local shares of $y$ from that of $x$. This operation will be denoted by  $[\![z]\!]_q\leftarrow[\![x]\!]_q-[\![y]\!]_q$.
  \item Multiplication by a constant ($z=cx$):
   Alice and Bob multiply their local shares of $x$ by $c$. This operation will be denoted by 
  $[\![z]\!]_q\leftarrow c[\![x]\!]_q$.
  \item Addition of a constant ($z=x+c$): Alice adds $c$ to her share $x$, while Bob keeps the same share of $x$. This operation will be denoted by $[\![z]\!]_q\leftarrow[\![x]\!]_q + c$.
\end{itemize}

It is a well-known fact that secure computations are impossible in a two-party setting unless additional (computational and/or setup) assumptions are in place. We work on the commodity-based cryptographic model, where a trusted initializer (TI) pre-distributes correlated randomness to the parties participating in the protocol. In particular, we make extensive use of pre-distributed multiplication triples. This \textit{multiplication triples} technique was originally proposed by Beaver \cite{beaver1997commodity} and is regularly used to enable very efficient solutions in the context of PPML (see e.g.~\cite{david2015efficient,AISec:CDNN15,mohassel2017secureml,guo2020secure}). 
The TI additionally generates random values in $\mathbb{Z}_q$ and delivers them to Alice so that she can use them to secret share her inputs. If Alice wants to secret share an input $x$, she picks an unused random value $r$ (note that Bob does not know $r$), and sends $c=x-r$ to Bob. 
Her share $x_A$ of $x$ is then set to $x_A=r$, while Bob's share $x_B$ is set to $x_B=c$. The secret sharing of Bob's inputs is done similarly using random values that the TI only delivers to him. After the setup phase, the TI is not involved in any other part of the execution and does not learn any data from the parties. In case a TI is not available or desirable, Alice and Bob can simulate the role of the TI, at the cost of additional pre-processing time and computational assumptions. 

We use the same protocol $\pmmul$ for secure (matrix) multiplication of secret shared values as in \cite{IEEETDSC:CDHK+17} and denote by $\pmul$ the protocol for the special case of multiplication of scalars. The notation $[\![Z]\!]_q \leftarrow[\![X]\!]_q \cdot [\![Y]\!]_q$ is used to denote the multiplication of two secret shared matrices $X$ and $Y$, and the notation $[\![v]\!]_q * [\![w]\!]_q$ is used to denote the element wise product of two secret shared vectors $v$ and $w$.

When working with fixed-point representations over $\mathbb{Z}_q$ with $a$ fractional bits, every multiplication generates an extra $a$ bits of unwanted fractional representation. Having a secure way to ``chop off'' the extra fractional bits generated by multiplication is a requirement to efficiently work with fixed-point secret shares. In the two-party scenario with shares in $\mathbb{Z}_{2^\lambda}$, it is possible to perform this truncation with a local probabilistic protocol -- hereafter, \ptrunc -- that with overwhelming probability in the security parameter $\lambda - (a + b)$ introduces an error of at most 1 in the least significant bit \cite{mohassel2017secureml}.  Since all operations to compute \ptrunc are local, the performance overhead of truncating all multiplication results with this method is essentially zero.

Often in MPC, there are problems that are best solved with integer arithmetic performed over $\mathbb{Z}_q$ (for a large $q$), while others (such as secure comparisons) are best solved over $\mathbb{Z}_2$. We work using a combination of these two techniques. Thus, it is necessary to be able to convert secret shares from one modulus to the other. For example, to determine if $\lss a \rss_q$ is equal to $\lss b \rss_q$, one would first convert $\lss a \rss_q$ and $\lss b \rss_q$ into bitwise sharings over $\mathbb{Z}_2$ and then confirm that each bit is identical using binary operations. The result, itself a bit shared over $\mathbb{Z}_2$, needs to be converted back to a sharing over $\mathbb{Z}_q$ for use in subsequent computations with integers. A quite efficient protocol for converting from $\mathbb{Z}_{2^\lambda}$ shares to bitwise sharings over $\mathbb{Z}_2$, denoted \pdecomp, can be found in \cite{idash}. \pdecomp is not fully described in this paper because novel methods for computing control flow logic are developed in the following section that do not require a full decomposition into $\mathbb{Z}_2$ shares. 

Regarding the opposite direction, the conversion from a bit shared over $\mathbb{Z}_2$ to a sharing of 0 or 1 over $\mathbb{Z}_q$, hereafter \ptoq \cite{NeurIPS2019}, is still necessary for our purposes. The intuition for \ptoq is that for a bit $\lss b \rss_2$ shared over $\mathbb{Z}_2$ between two parties, there are four possible secret shares of which three are valid $\mathbb{Z}_q$ sharings and one is not. If $\lss b \rss_2 = b_0 + b_1 \mod 2= 1$, then $(1,0)$ and $(0,1)$ are the only possible sharings. Similary, if $\lss b \rss_2 = b_0 + b_1 \mod 2 = 0$, then $(0,0)$ and $(1,1)$ are the only possible sharings. In the case of secret shares $(0, 0)$, $(0, 1)$, and $(1, 0)$, it holds automatically that $b_0 + b_1 \mod  2 = b_0 + b_1 \mod q$ for all $q > 2$. However, the problematic sharing $(1,1)$ -- which encodes $0$ -- sums to $2$ when regarded as a $\mathbb{Z}_q$ sharing. Hence, the problem of converting from $\mathbb{Z}_2$ to $\mathbb{Z}_q$ is reduced to mapping a secret sharing of $2$ to a secret sharing of $0$ as described in Protocol \ref{prot:2toq}.\\

\begin{procedure}
    \SetKwInOut{Input}{Input}
    \SetKwInOut{Output}{Output}

    \Input{ $\lss b \rss_2 \; := \; (b_0, \; b_1)$ }
    \Output{$\lss b \rss_q$ }
    
    Alice creates the sharing $\lss b_0 \rss_q = (b_0, 0 )$ \\
    
    Bob creates the sharing $\lss b_1 \rss_q = (0, b_1)$ \\
    
    $\lss b \rss_q \leftarrow \lss b_0 \rss_q + \lss b_1 \rss_q - 2 \cdot \lss b_0 \rss_q \cdot \lss b_1 \rss_q$
    
    \KwRet{$\lss b \rss_q$}
    
    \caption{Secure Protocol() $\ptoq$ converts a secret bit from a $\mathbb{Z}_2$ sharing to a $\mathbb{Z}_q$ sharing.}
    \label{prot:2toq}
\end{procedure}

\noindent
\textbf{Protocol for secure DT training.}
Each internal node in a DT tests the value of a particular feature and branches out accordingly, while each leaf node contains a class label. In the clear, training of a DT is done by growing the tree from the root to the leaf nodes in a recursive manner. For each internal node, the feature is selected that splits the set of training instances that have reached that node in subsets that are as homogeneous as possible regarding the class label value. MPC protocols for secure training of DTs commonly use the Gini impurity to this end (the lower the impurity, the higher the homogeneity). As a sub-protocol for performing the secure training of a DT with categorical data we use protocol \pSIDT that is a slightly modified version of the protocol SID3T of De Hoogh et al. \cite{de2014practical}. The differences are the following:

\begin{itemize}[leftmargin=*,topsep=0pt]
    \item De Hoogh et al.~\cite{de2014practical} used secret shares in a prime field because their multiplication protocol required the existence of modular inverses. We instead perform multiplications using multiplication triples and can work with shares in $\mathbb{Z}_{2^\lambda}$.
    \item For stopping criteria, De Hoogh et al.~\cite{de2014practical} used: (1) no features remain in the training set in the node at hand, (2) all remaining instances in the node have the same class label, and (3) the number of remaining instances in the node is less than a cutoff threshold. We use only (2) and (3), and additionally work with a pre-specified maximum allowed tree depth.
    \item Their solution grows the DT recursively and leaks the shape of the tree. Our version grows DT's iteratively (by depth, up to a max depth) and adds dummy nodes where needed to hide true path lengths. We call the last non-dummy node on each path a \textit{classifying node}. We keep track of such nodes by cascading a secret shared bit representing whether or not an early stopping condition was already reached on the path during training.
%    and tracks where the base cases have been reached, cascading a secret shared bit representing whether or not any parent node has reached an early stopping condition, thereby preventing multiple nodes from classifying within a single path. The final tree is a complete tree where any unique path contains only one classifying node.
    \item For each leaf node, De Hoogh et al. \cite{de2014practical}'s protocol returns a secret-shared one-hot-encoded vector denoting the class label. In contrast, for each classifying node, our algorithm returns secret shared frequencies of each of the class labels in the subset of training instances that have reached that node. Such secret shared frequency values are inexpensive to compute because additions can be done locally by the computing parties, and the frequencies allow for weighted aggregation of class label votes of trees in an ensemble, leading to more accurate classifications. 
\end{itemize}

\noindent
The DT is output as an array of secret shared one-hot-encodings of the split feature at each node, in addition to the value(s) to be compared against.

%%%%%%%%%%%%%%%%%%%%%%%%%%%%%%%%%%%%%%
%%
%%   3. Secure Comparison
%%
%%%%%%%%%%%%%%%%%%%%%%%%%%%%%%%%%%%%%

\section{Secure Comparison Protocol}\label{SEC:COMP}

Secure comparison of integers is a well-studied problem in this domain. Specifically, many solutions exist to compute the output of non-linear functions of the form
\[ \lss x \rss_q \geq^? \lss y \rss_q \; : \; \lss 1 \rss_2 \;\; \mathbf{else} \;\; \lss 0 \rss_2, \]
\[ \lss x \rss_q =^? \lss y \rss_q \; : \; \lss 1 \rss_2 \;\; \mathbf{else} \;\; \lss 0 \rss_2. \]
 
The most common solutions involve first converting the inputs to their corresponding bitwise sharings $\lss x \rss_q \rightarrow \lss x_\lambda \rss_2 \cdots \lss x_1 \rss_2$ and $\lss y \rss_q \rightarrow \lss y_\lambda \rss_2 \cdots \lss y_1 \rss_2$ for $\lambda \geq \lceil \log(q) \rceil$ such that $\sum x_i 2^i = x$ and $\sum y_i 2^i = y$. Afterward, a series of bitwise operations are carried out to determine which of the two bit strings is greater than or equal to the other. The efficiency of a given protocol depends on the constraints of the secure computational environment for which it is designed. We direct the reader to \cite{DBLP:journals/ieicet/AttrapadungHKMS19}
for a comprehensive discussion of comparison protocol design. For the computational environment used in this work, (i.e. 2PC, semi-honest security setting, linear secret sharing modulo $2^\lambda$), the most similar protocol to our work is proposed by Bogdanov et al. for the Sharemind framework \cite{10.1007/978-3-540-88313-5_13}, though this framework is designed for 3PC.
Our work and that of \cite{10.1007/978-3-540-88313-5_13} rely on observations that hold for two's complement representations in $\mathbb{Z}_{2^\lambda}$, and work for numbers $x$ and $y$ such that $|x-y| < 2^{\lambda-1}$ holds, which can be easily enforced by only using a sub-range which is less than half of the available range $2^\lambda-1$. This is a relatively weak limitation because any desired range can be injected into a larger integer ring. However, it fails for applications that rely on other MPC protocols for which the existence of modular inverses must be guaranteed.

The key insight we use to compute $x \geq^? y$ securely is the following: for $x, \; y$ in two's complement form and such that $|x-y| < 2^{\lambda-1}$, we have that $y >x \iff 0 > x - y \iff \mathsf{MSB}(x-y) = 1$, where $\mathsf{MSB}(\cdot)$ denotes the \textit{most significant bit} of a value. %Then, 
Hence $x \geq y \iff \mathsf{MSB}(x-y) = 0$.
Similar logic can be used to derive an equality check $x =^? y$:
\begin{equation*}
    \begin{split}
        x = y &\iff (x \geq y) \land (y \geq x) \\
        &\iff \mathsf{MSB}(x-y)=\mathsf{MSB}(y-x)=0\\
    \end{split}
\end{equation*}

Note that due to the two's complement format, it is not possible to have $\mathsf{MSB}(x-y)=\mathsf{MSB}(y-x)=1$.

The above shows that the efficiency of computing $x \geq^? y$ and $x =^? y$ is limited only by our ability to extract the most significant bit of a secret-shared value. Bogdanov et al. use a recursive carry look-ahead construction to decompose the difference $x-y$ into its bitwise sharing. Other solutions designed for sharing over prime fields use a similar approach but take advantage of the fact that $\mathsf{MSB}(x) = \mathsf{LSB}(2x \mod  q)$ for odd $q$ \cite{10.1007/978-3-540-71677-8_23}, where $\mathsf{LSB}(\cdot)$ denotes the least significant bit. 

In this section, we propose protocols for comparison and equality tests over $\mathbb{Z}_{2^\lambda}$ -- \pgeq and \peq, respectively -- that circumvent a full-blown bit decomposition by extracting only the most significant bit -- causing a significant reduction to the total data transfer. Our approach is based on a modification of the optimized bit decomposition protocol $\pdecompopt$ presented in \cite{idash}. 

Note that when working with a two's complement fixed-point representation over $\mathbb{Z}_{2^\lambda}$, all bits outside of an injected ring $\mathbb{Z}_{2^{a+b}}$, where $a$ is the number of fractional bits and $b$ is the number of integer bits, are equivalent to the most significant bit. It follows in this case that extracting the $(a+b+1)$-th bit is equivalent to extracting the most significant bit which further reduces the depth of the arithmetic circuit. 

\begin{procedure}
    \SetKwInOut{Input}{Input}
    \SetKwInOut{Output}{Output}

    \Input{ $\lss x\rss,\; \lss y \rss$ such that $|x-y| < 2^{\lambda-1}$, $\alpha $ := the lowest bit position guaranteed to be equal to the MSB.}
    \Output{$\lss x \rss \geq^? \lss y \rss \; : \; [\![1]\!]_{2} \;\mathbf{else} \; [\![0]\!]_{2}$}
    Let $ \lss \mathsf{diff} \rss \leftarrow \lss x \rss - \lss y \rss $ \\
    Let $ \lss \mathsf{MSB} \rss_2 \leftarrow \pbtx( \lss \mathsf{diff} \rss, \alpha )$\\

    \KwRet{$1 \oplus [\![ \mathsf{MSB} ]\!]_2$}
    \caption{Secure Protocol() $\pgeq$ computes the integer comparison.}
    \label{prot:geq}
\end{procedure}

\begin{procedure}
    \SetKwInOut{Input}{Input}
    \SetKwInOut{Output}{Output}

    \Input{ $\lss x\rss,\; \lss y \rss$ such that $|x-y| < 2^{\lambda-1}$, $\alpha $ := the lowest bit position guaranteed to be equal to the MSB.}
    \Output{$\lss x \rss =^? \lss y \rss \; : \; [\![1]\!]_{2} \;\mathbf{else} \; [\![0]\!]_{2}$}
    
    Let $ \lss \mathsf{d}_1 \rss \leftarrow \lss x \rss - \lss y \rss $ and $ \lss \mathsf{d}_2 \rss \leftarrow \lss y \rss - \lss x \rss $ \\

    Let $ \lss \mathsf{MSB}_1 \rss_2 \leftarrow \pbtx( \lss \mathsf{d}_1 \rss, \alpha )$ 
    and $ \lss \mathsf{MSB}_2 \rss_2 \leftarrow \pbtx( \lss \mathsf{d}_2 \rss, \alpha )$// run  in parallel \\

    \KwRet{$1 \oplus \lss \mathsf{MSB}_1 \rss_2 \oplus \lss \mathsf{MSB}_2 \rss_2$}
    \caption{Secure Protocol() $\peq$ computes the integer equality test.}
    \label{prot:eq}
\end{procedure}

The two-party protocol $\pdecompopt$ for performing a full decomposition of a $\mathbb{Z}_{2^\lambda}$-shared secret into bitwise sharings over $\mathbb{Z}_2$ is, to our knowledge, the most efficient in the literature. It is based on a \textit{matrix composition network} that computes the difference between each bitwise sum of two secret shares and the corresponding ``actual'' bit of the secret value in $\mathbb{Z}_{2^\lambda}$. See \cite{idash} for a complete description. An important aspect of this approach is that computing the difference for the $\alpha$-th bit depends only on $\lceil \log(\alpha - 1) \rceil$ rounds of matrix composition, where each matrix composition requires 4 bits of data transfer. In addition, the difference for the $\alpha$-th bit is independent of all results for lower order bits.

In Protocols \ref{prot:geq} and \ref{prot:eq}, we make use of a straightforward modification of $\pdecompopt$ for extracting the $\alpha$-th bit from a $\mathbb{Z}_{2^\lambda}$-shared secret between two parties, hereafter \pbtx (see Appendix \ref{sec:btx} for details). The total number of communication rounds to extract the $\alpha$-th bit is $\lceil \log(\alpha-1) \rceil + 1$ with total data transfer of $ 2(\alpha-1) + 4 \lceil \log(\alpha-1) \rceil$ bits.    

%%%%%%%%%%%%%%%%%%%%%%%%%%%%%%%%%%%%%%%%%%%%%%%%%%%%%%%
%
%      3. SECURE DISCRETIZATION
%
%%%%%%%%%%%%%%%%%%%%%%%%%%%%%%%%%%%%%%%%%%%%%%%%%%%%%%%
\section{Secure Discretization}
Discretization or ``binning'' is a common form of data preprocessing, aimed at grouping continuous or numerical values into a smaller number of bins (buckets). In a data set with information about social media users, the feature \textit{age} could for instance be discretized into the bins 0-24, 25-34, 35-49, 50+. The threshold values for the bins are typically derived from the data in an unsupervised manner. We use \textit{equal-width binning}, which means that the range of feature values is divided into a predefined number of bins of the same width.

Let $D$ be a vector containing the original feature values (e.g.~the ages of all the users in the data set). The range of $D$ is bounded by the smallest and the largest value occurring in $D$, i.e.~$\min(D)$ and $\max(D)$. To divide this range into $p$ bins of the same width, thresholds need to be placed at
\begin{equation}\label{eq:thresholds}
h_i = \min(D) + i \cdot \frac{\max(D)-\min(D)}{p} 
\end{equation}
for $i = 1,\ldots,p-1$. The challenge is that neither Alice nor Bob may have direct access to $\min(D)$ and $\max(D)$ because they each may only have shares of the values of $D$ (see Figure~\ref{fig:diagram}). This means that they need to jointly execute a secure protocol, hereafter $\pminmax$, for computing the minimum and the maximum values of $D$. 

%%%%%%%%%%%%%%%%%%%%%%%%%%%%%%%%%%%%%%%%%%
%% Linear p_minmax                    
%%%%%%%%%%%%%%%%%%%%%%%%%%%%%%%%%%%%%%%%%

\begin{procedure}[t]
    \SetKwInOut{Input}{Input}
    \SetKwInOut{Output}{Output}
    \Input{$\lss D \rss$, number $n$ of elements in $\lss D \rss$}
    \Output{ $\lss d_\mathsf{min} \rss$,  $\lss d_\mathsf{max} \rss$} 
    
    Let $\lss \geq^? \rss \leftarrow \ptoq( \pgeq ( \lss d_2 \rss, \lss d_1 \rss ) )$ \\
    
   Let $\lss d_\mathsf{min} \rss \leftarrow \lss \geq^? \rss \cdot \lss d_1 \rss + (1 - \lss \geq^? \rss) \cdot \lss d_2 \rss $ \\
   Let $\lss d_\mathsf{max} \rss \leftarrow \lss \geq^? \rss \cdot \lss d_2 \rss + (1 - \lss \geq^? \rss) \cdot \lss d_1 \rss $ \\

   \For{$i\gets 3$ \KwTo $n$} {
   
    $ \lss \geq^?_\mathsf{min} \rss \leftarrow \ptoq( \pgeq( \lss d_i \rss, \lss d_\mathsf{min} \rss ) ) $ \\

   $ \lss \geq^?_\mathsf{max} \rss \leftarrow \ptoq( \pgeq( \lss d_i \rss, \lss d_\mathsf{max} \rss ) ) $ \\ 
   
   $ \lss d_\mathsf{min} \rss \leftarrow \lss \geq^?_\mathsf{min} \rss \cdot \lss d_\mathsf{min} \rss + (1 - \lss \geq^?_\mathsf{min} \rss ) \cdot \lss d_i \rss $ \\
   
   $ \lss d_\mathsf{max} \rss \leftarrow \lss \geq^?_\mathsf{max} \rss \cdot \lss d_i \rss + (1 - \lss \geq^?_\mathsf{max} \rss ) \cdot \lss d_\mathsf{max} \rss $ \\
   }
    
    \KwRet{$ \lss d_\mathsf{min} \rss, \; \lss d_\mathsf{max} \rss  $}

    \caption{Secure Min/Max-Finding Protocol() \pminmax \label{prot:minmax}}
\end{procedure}

To compute secret sharings of min($D$) and max($D$) without revealing any information about $D$, a secure protocol can be formulated similarly to a naive sequential search solution in the clear. That is, start by comparing the first and second elements of $D$ to determine an initial estimate of the max and min. Next, iterate through all remaining elements and adjust the max and min estimates when a new largest or smallest element is found. After the $n$-th element of $D$ is checked, the estimates are guaranteed to be the global min and max. The only necessary adaptation for this algorithm to act as an oblivious protocol is to require that the comparisons between the current estimates and each new element of $D$ are performed with $\pgeq$ and that the reassignments are handled with multiplication rather than control flow logic. For example, the comparison based branch operation ``$\textbf{if  } a \geq b \textbf{  then  } b = a$'' can be rephrased as 
\begin{equation}
\begin{split}
        \lss c \rss \gets \ptoq(\pgeq(\lss a \rss, \lss b \rss)) \\
        \lss b \rss \gets \lss c \rss \cdot \lss a \rss + (1- \lss c \rss) \cdot \lss b \rss \\
\end{split}
\end{equation}
where $c$ is 1 or 0, depending on the outcome of the comparison $a \geq b$. This form of conditional assignment doesn't allow Alice nor Bob to learn anything about which branch of the control flow sequence was followed to arrive at the outcome. An additional detail is that because \pgeq returns secret shares over $\mathbb{Z}_2$, the result must first be converted to a ring representation with \ptoq before the multiplication can be carried out.

\begin{figure}
    \includegraphics[width=\columnwidth]{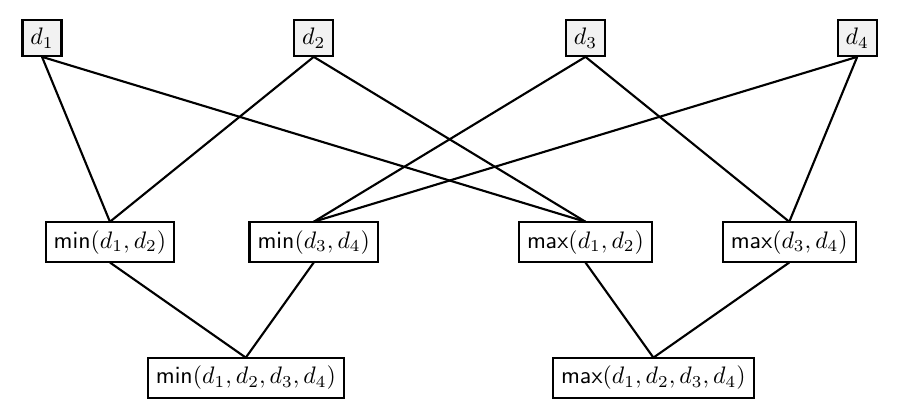}
    \caption[width=\columnwidth]{Optimized $\pminmax$: 
    An example circuit to compute \pminmax on an input vector of size $n=4$}
    \label{fig:minmax}
\end{figure}

Protocol \pminmax has a linear number of communication rounds when carried out in the naive formulation that is described in Protocol \ref{prot:minmax}. However, it can be improved straightforwardly with the same optimisation technique used for securely computing the repeated product over a vector of values in which pairwise products are taken until only one value remains \cite{IEEETDSC:CDHK+17}. The protocol \pminmax is analogous to repeated multiplication because both multiplication and the max/min functions are associative. So, the sequence of many repeated applications can be altered to reduce the total number of consecutive, mutually dependent applications. The basic observation is that the global minimum of $D$ is contained in the set of all pairwise minima of $D$. Moreover, if the minimum is computed between each pair of entries, and this process is repeated until only one pair remains, the result of the final pairwise comparison is min($D$). The same principle extends to finding the global maximum. 
See Figure~\ref{fig:minmax} for an illustration.

As a result, \pminmax can be computed in a circuit of depth $\lceil \log(n) \rceil$, where at each layer there are $\lceil \log(a + b) \rceil + 1$ rounds of communication, where $a$, $b$ are the number of fractional and integer bits, respectively, used in the representation.

At the end of this protocol, Alice and Bob have secret sharings of $\min(D)$ and $\max(D)$, which they can then securely combine to compute secret sharings of each of the $h_i$ thresholds in Equation (\ref{eq:thresholds}). For example, if $\min(D) = 0$, $\max(D)=150$, and $p=6$, then Alice and Bob would compute secret shares of the thresholds $25, 50, 75, 100,$ and $125$. We assume that $p$ is publicly known, i.e.~Alice and Bob know how many bins need to be created. As explained above, they may however not know the value of $\min(D)$ or of $\max(D)$ in the clear, and our discretization protocol does not leak this value.

After they have computed shares of the $h_i$ thresholds, Alice and Bob can map each value $d_j$ from $D$ into its correct bin number $bin(d_j)$ by executing protocol $\pdisc$ (described in Protocol \ref{prot:disc}). Note that Alice and Bob each only have a share of $d_j$, and a share of each threshold $h_i$. They could run the secure comparison protocol $\pgeq(d_j,h_i)$ for each of the threshold values $h_i$ and count in how many cases the comparison yielded a ``true'' response. For example, for value $d_j=80$ and thresholds $25, 50, 75, 100,125$, the first 3 comparison tests with $\pgeq$
would result in secret shares of $1$, while the remaining 2 tests would result in secret shares of $0$. Adding those up securely, Alice and Bob would derive shares of $3$, which is the correct bin number of $d_j=80$, assuming that we start counting bins at 0.

However, in order to expedite the process of converting this discretized data into the required format used by the  decision tree learning protocols that follow, we choose to output a one-hot-encoding of $bin(d_j)$ where the $bin(d_j)$-th bit position (with lowest order on the left) is a secret sharing of 1 and all other bits are secret sharings of 0. Building from the previous example where $p=6$ and $bin(d_j) = 3$, \pdisc outputs secret shares of the vector $(0, 0, 0, 1, 0, 0)$.

%%%% PROTOCOL pi_DISC %%%%%%%%%%%%%%%%%%%%%%%

\begin{procedure}[t]
    \SetKwInOut{Input}{Input}
    \SetKwInOut{Output}{Output}

    \Input{$[\![D]\!]$, number $n$ of elements in $[\![D]\!]$, public number of buckets $p$}
    \Output{$[\![D']\!]_2$ := one-hot-encoding of the bucket membership of each $d \in D$} 
    
    The parties call \pminmax on $\lss D \rss$ and receive $\lss d_\mathsf{min} \rss$,  $\lss d_\mathsf{max} \rss$ \\
    
    $\lss d_\mathsf{range} \rss \leftarrow \lss d_\mathsf{max} \rss - \lss d_\mathsf{min} \rss$ \\
    
    \For{$i\gets 1$ \KwTo $p-1$} {
        $\lss h_i \rss \leftarrow \lss d_\mathsf{min} \rss + \ptrunc( \frac{i}{p} \cdot \lss d_\mathsf{range} \rss)$ \\
    }
    % \For{$j\gets 1$ \KwTo $n$} {
    %     Let $\lss d'_j \rss \leftarrow \sum\limits_{i=1}^{p-1}\ptoq( \pgeq ([\![d_j]\!], [\![h_i]\!]))$ \\
    % }
    \For{$i\gets 1$ \KwTo $n$} {
        
       $\lss e_j \rss_2 \gets \pgeq(\lss d_i \rss, \lss h_j \rss) $ for $j \in 1, \ldots, p-1$\\ 
    
        $\lss d'_{i,(0)} \rss_2 \gets 1 - \lss e_1 \rss_2 $\\
            
        $\lss d'_{i,(j)} \rss_2 \gets \lss e_j \rss_2 \cdot (1 - \lss e_{j + 1} \rss_2)$ for $j \in 1, \ldots, p-2$ \\
        $\lss d'_{i,(p-1)} \rss_2 \gets \lss e_{p-1} \rss_2$ \\
    }
    
    \KwRet{$[\![D']\!]_2$}
 
    \caption{Secure Equal-Width Discretization Protocol() \pdisc  \label{prot:disc}}
 
\end{procedure}

This conversion to one-hot-encoding is carried out by noting that any value compared against a set of $p-1$ increasing thresholds will return $0 \leq k \leq p-1$ true results followed by $p-k-1$ false results. The position of the 1 in the one-hot-encoding vector is determined by the position of the first false result. In the protocol that follows, the notation $d'_{i,(j)}$ means the j-th bit of the one-hot-encoding vector for $d_i$. 

Protocol \pdisc (detailed in Protocol \ref{prot:disc}) adds only additional $\lceil \log(a+b)\rceil + 2$ communication rounds after \pminmax, where $a, b$ are the fractional and integer precision of the injected fixed point subring, respectively. All calls to \pgeq are mutually independent (Line 7), so they are computed in a single batch in $\lceil \log(a+b)\rceil + 1$ rounds. Similarly, all products $e_j \cdot (1-e_{j+1})$ (Line 9) are mutually independent and require one round. Computing the range and thresholds $\lss h_i \rss$ (Line 2-5) is entirely comprised of local operations as all constants $i/p$ are public and \ptrunc is a local protocol in the 2PC scenario.

%%%%%%%%%%%%%%%%%%%%%%%%%%%%%%%%%%%%%%%%%%%%%%%%%%%%%%%
%
%      4. SECURE RF
%
%%%%%%%%%%%%%%%%%%%%%%%%%%%%%%%%%%%%%%%%%%%%%%%%%%%%%%%
\section{Secure Random Forest Training}\label{SEC:RF}
DT ensembles gained much popularity during the 2000s and have remained state-of-the-art methods for many classification tasks to date. A DT ensemble consists of a set of DTs that each infer a class label for a new instance; the final label is determined through (weighted) majority voting. DT ensembles differ from each other in the way they are trained. A successful approach, known as random forest (RF), combines \textit{bagging} and \textit{subspace sampling} to make the DTs in the ensemble sufficiently different from each other \cite{breiman2001random}.

\textit{Bagging} refers to the fact that, given a training data set $S$ with $n$ examples, for each DT a bootstrap replica of $S$ is created by sampling $n$ times with replacement from the data set $S$. In this paper, we present a slight adaptation of the traditional RF training algorithm in which we sample only $s \leq n$ times. Working with a smaller data set allows to train a RF more efficiently in case $n$ is very large. 
%In Section \ref{sec:results} we investigate the effect of this on the accuracy of the trained models.
\textit{Subspace sampling} refers to the fact that for each DT, only $k < f$ randomly selected features of the original feature set are retained.
%Random forest is an ensemble algorithm, which is well-known for its ``bagging'' idea and random feature selection technique. Let $S$ be a training set which contains $n$ instances (each with $f$ features and one target class with $c$ possible classifications). For a random forest model with $m$ trees, the algorithm will sample the training set with replacement $s$ times independently before training each tree. To inject more randomness, the algorithm also randomly selects $k$ features to be used in each tree. The algorithm then train $m$ shallow trees in parallel using these data sets. 
The resulting RF training algorithm -- in the clear -- is presented in Algorithm \ref{prot:rfclear}. Note that the use of ID3 \cite{quinlan1986induction} on Line 4 indicates that the feature values are assumed to be categorical; otherwise one would typically use C4.5 \cite{quinlan2014c4} instead.

\begin{algorithm}[t]
    \SetKwInOut{Input}{Input}
    \SetKwInOut{Output}{Output}
    \Input{A set $S$ with $n$ training samples (each sample has $f$ features and one class label),
    %target class with $c$ possible classifications)
    the number $m$ of trees in an ensemble, the number of features $k$ used in each tree, the number of samples $s$ used in each tree, the depth $d$ of each tree.}
    \Output{An ensemble of trees $T = t_{1}, t_{2}, \ldots, t_{m}$} \For{$j\gets 1$ \KwTo $m$}
    {
        Randomly select $k$ of the $f$ features of $S$.\\
        Randomly select with replacement $s$ samples $S' = i_{1}, i_{2}, \ldots, i_{s}$  among all samples of $S$ after features are selected.\\
       Train a decision tree $t_{i}$ of depth $d$ on $S'$ using ID3.\\
   }
    \KwRet{$T = t_{1}, t_{2}, \ldots, t_{m}$}
    \caption{Algorithm for Training a Random Forest Classifier.}\label{prot:rfclear}
\end{algorithm}

Our approach for training a RF in a privacy-preserving manner over data with continuous feature values is to first discretize the feature values using $\pdisc$. 
%and then run Protocol $\pRF$ (see Protocol \ref{PROT:RF}). 
Next we need techniques for randomly selecting features and samples in a secure manner, similar to Line 2 and 3 in Algorithm \ref{prot:rfclear}. Finally, we train DTs as in Line 4, using the secure protocol $\pSIDT$ adapted from de Hoogh et al.~\cite{de2014practical}.
$\pSIDT$ assumes that the input data is presented in a one-hot-encoded (OHE) format, because that allows for efficient calculation of the Gini index which is needed to select split features while training a DT. That is the reason why we designed $\pdisc$ to already output  one-hot-encodings. Note that after a discretization of all feature values of the data set $S$, using $p$ bins per feature, the secret shared data set $[\![S_{\mathsf{disc}}]\!]_2$ consists of 
a matrix $S_{\mathsf{disc}}$ of size $n \times f \cdot p$ containing the one-hot-encodings. For oblivious selection of $k$ features we use a $f \cdot p \times k \cdot p$ selection matrix $FS$ that is generated by the TI and secret shared with the parties. In $FS$, identity submatrices of size $p \times p$ are used to indicate that a feature is selected, and the remaining positions are filled with $p \times p$ submatrices of zeroes. Since each feature is selected at most once, no identity submatrices are aligned horizontally or vertically. Note that $m$ such matrices are randomly populated in this manner by the TI, with $m$ the total number of DTs in the RF. In order to extract the desired features, the secure protocol for random feature selection %$\prfs$ 
calls the secure matrix multiplication protocol $\pmmul$ to
multiply the OHE-style data set $[\![S_{\mathsf{disc}}]\!]_2$  with a feature selection matrix $FS$ that is secret shared by the TI. For example, if $f=3$, $p=3$ and $k=2$, the matrix
\begin{equation*}
{
FS = 
\begin{pmatrix}
1 & 0 & 0 & 0 & 0 & 0 \\
0 & 1 & 0 & 0 & 0 & 0 \\
0 & 0 & 1 & 0 & 0 & 0 \\
0 & 0 & 0 & 0 & 0 & 0 \\
0 & 0 & 0 & 0 & 0 & 0 \\
0 & 0 & 0 & 0 & 0 & 0 \\
0 & 0 & 0 & 1 & 0 & 0 \\
0 & 0 & 0 & 0 & 1 & 0 \\
0 & 0 & 0 & 0 & 0 & 1 \\
\end{pmatrix}
}
\end{equation*}
will retain the first 3 and the last 3 columns of $S_{\mathsf{disc}}$, thereby effectively selecting the one-hot-encodings corresponding to the first and third features of $S$.

The procedure used to sample the $s$ instances with replacement is similarly done by a multiplication with a $s \times n$ matrix $SS$ that is secret shared by the TI. The only difference is that the identity matrices used for selecting can be aligned, as the choice is with replacement. 

%%%% ALGORITHM %%%%%%%%%%%%%%%%%%%%%%%
\begin{procedure}[t]
    \SetKwInOut{Input}{Input}
    \SetKwInOut{Output}{Output}

    \Input{A secret shared set $[\![S]\!]$ with $n$ training samples (each sample has $f$ features), the number $m$ of trees in an ensemble, the number of buckets $p$ for each feature, the number of features $k$ used in each tree, the number of samples $s$ used in each tree, the depth $d$ of each tree.}
    \Output{A random forest model $[\![RF]\!]=[\![t_{1}]\!], \ldots, [\![t_{m}]\!]$.} 
     (Offline Phase) The TI generates and secret shares $m$ 0/1 valued feature selection matrices
$[\![FS^{(1)}]\!]_2, \ldots, [\![FS^{(m)}]\!]_2$ of size $f \cdot p \times k \cdot p$, and $m$ 0/1 valued sample selection
matrices $[\![SS^{(1)}]\!]_2, \ldots, [\![SS^{(m)}]\!]_2$ of size $s \times n$\\
    Discretize each feature of $[\![S]\!]$ into $p$ bins using protocol $\pdisc$ to get 
    $[\![S_{\mathsf{disc}}]\!]_2$\\
    \For{$i\gets 1$ \KwTo $m$}{
        $[\![S_{\mathsf{FS}}]\!]_2 \leftarrow  [\![S_{\mathsf{disc}}]\!]_2 \cdot [\![FS^{(i)}]\!]_2$. \\
        $[\![S_{\mathsf{SR}}]\!]_2 \leftarrow  [\![SS^{(i)}]\!]_2 \cdot [\![S_{\mathsf{FS}}]\!]_2$. \\
        Use $\pSIDT$ to securely train a decision tree $[\![t_{i}]\!]$ of depth $d$ with the data set $[\![S_{\mathsf{SR}}]\!]_2$.
    }
    \KwRet{$[\![RF]\!]=[\![t_{1}]\!], \ldots, [\![t_{m}]\!]$.}
    
    \caption{Secure Protocol() $\pRF$ for Training a Random Forest with Continuous Data. \label{PROT:RF}}
\end{procedure}

Protocol $\pRF$ for secure training RFs is described in Protocol \ref{PROT:RF}. The loop in $\pRF$ can be executed in parallel. The original ID3 protocol will not grow a tree to pre-specified depth $d$ if an early termination condition is satisfied (e.g.~all training examples in a branch have the same class label). We have modified it in $\pSIDT$ such that the secure trees will always grow to depth $d$ by adding dummy nodes where necessary. If a node satisfies an early termination condition, then the node will override the classifications of all of its child nodes (dummy nodes) in an oblivious manner. This has the security advantage of not revealing the true depth of each sub-tree, while concealing which nodes actually classify.

%%%%%%%%%%%%%%%%%%%%%%%%%%%%%%%%%%%%%%%%%%%%%%%%%%%%%%%
%
%      5. SECURE XT
%
%%%%%%%%%%%%%%%%%%%%%%%%%%%%%%%%%%%%%%%%%%%%%%%%%%%%%%%

\section{Secure Extra Trees Training}
\label{SEC:PROT}
Besides the RF training algorithm from the previous section, several other successful algorithms exist for training ensembles of decision trees. One of these algorithms, used to train so-called ``Extremely Randomized Trees'', or ``extra-trees'' (XT) for short, was developed specifically for data with numerical features \cite{geurts2006extremely}. In addition to randomly selecting subsets of features during the tree construction process, the XT training algorithm also randomly selects a threshold $\alpha_j$ for each feature $a_j$ to effectively turn the numerical feature $a_j$ into a binary feature, based on whether the feature value is greater than or equal to $\alpha_j$, or not.

\begin{algorithm}[t]
    \SetKwInOut{Input}{Input}
    \SetKwInOut{Output}{Output}

    \Input{A training set $S$ with continuous data and $n$ samples (each sample has $f$ features), the number $k$ of features to consider in each tree, the number $m$ of trees in the ensemble, the depth $d$ of each tree.}
    \Output{An ensemble of trees $XT=t_{1}, \ldots, t_{m}$.} 
    
    For each feature $j$, find its minimum and maximum values, $\text{min}_j$ and $\text{max}_j$. \\
    \For{$i\gets 1$ \KwTo $m$}{
       Select $k$ random indices $j_1, \ldots, j_k \in \{1,\ldots,f\}$.\\
        \For{$\ell \gets 1$ \KwTo $k$}{
            For a uniformly random $r \in (0,1)$, $\alpha_{\ell} \gets r\cdot(\text{max}_{j_\ell} - \text{min}_{j_\ell}) + \text{min}_{j_\ell}$. \\
            \For{$s\gets 1$ \KwTo $n$}{
                \eIf{$S[s,j_\ell] \geq \alpha_{\ell}$}{
                    $S'[s,\ell] \leftarrow 1$
                }{
                    $S'[s,\ell] \leftarrow 0$
                }
            }
        }
        Train a decision tree $t_i$ of depth $d$ on $S'$ using ID3.
        %Let $t_i$ be the ID3 decision tree of depth $d$ trained with the data set $D$.
    }
    \KwRet{$XT=t_{1}, \ldots, t_{m}$.}
    
    \caption{Algorithm for Training an Extra-Trees Classifier. 
    \label{FIGEXTRATREESALG}}
\end{algorithm}

An algorithm for training an XT classifier in the clear, adapted from \cite{geurts2006extremely}, is presented in Algorithm \ref{FIGEXTRATREESALG}. It constructs an ensemble of $m$ decision trees. As before, $S$ can be thought of as an $n \times f$ matrix in which the rows correspond to instances and the columns to features. As in Section \ref{SEC:RF}, each decision tree is trained over a randomly chosen subset of $k$ of the $f$ available features. Moreover, each continuous feature $a_j$ is binarized by choosing a random threshold $\alpha_j$ in the range of possible values of $a_j$, and replacing the feature value by 1 if it meets the threshold, and 0 otherwise. Each such binarization is specific for a particular feature in a particular tree of the ensemble; for another tree in the ensemble the same feature $a_j$ might be reused with a different random binarization. Note that all loops in the algorithm can be executed in parallel. Algorithm \ref{FIGEXTRATREESALG} differs from the original extremely randomized trees algorithm \cite{geurts2006extremely} in the sense that in the latter the random choices for the features and the random choices for the cut-off points are made for each node in each decision tree, whereas in Algorithm \ref{FIGEXTRATREESALG} they are made once for each decision tree. We have observed that the difference between the two approaches can have some detrimental effect on the accuracy. To remedy this, we can sample the subspace of $S$ with replacement, potentially choosing multiple splits per feature. So for example, if we have a data set where $f = 30$, we can let $k = 60$, thus increasing the diversity of the data set. In our tests, this approach seems to close the gap in accuracy caused by discretizing per tree instead of per node.

\begin{procedure}[t]
    \SetKwInOut{Input}{Input}
    \SetKwInOut{Output}{Output}

    \Input{A secret shared training set $[\![S]\!]$ with continuous data and $n$ samples (each sample has $f$ features), the number $k$ of features to consider in each tree, the number $m$ of trees in the ensemble, the depth $d$ of each tree.}
    \Output{A secret shared ensemble of trees $[\![XT]\!]=[\![t_{1}]\!], \ldots, [\![t_{m}]\!]$.} 
    
    (Offline Phase) The TI secret shares $m$ random  0/1-valued feature selection matrices $\lss FS^{(1)} \rss, \ldots, \lss FS^{(m)} \rss$ of size $f\times k$, where each column contains a single 1, and no rows have more than a single 1. The TI also distributes $k \cdot m$ uniformly random ratios $\lss r^{(1)} \rss, \ldots, \allowbreak \lss r^{(k \cdot m)} \rss \allowbreak \in [1, 2^a - 1]$ (which approximates $r \in (0,1)$ in $\mathbb{R}$).\\
    
    Compute the vectors $\lss min \rss$ and $\lss max \rss$, by using $(\lss min_j \rss, \lss max_j \rss) \gets \pminmax(\lss S_j \rss,n)$ for each column $S_j$ ($j=1,\ldots,f$) of $S$. \\
    \For{$i\gets 1$ \KwTo $m$}{
       $[\![S_{\mathsf{FS}}]\!] \gets [\![S]\!] \cdot [\![FS^{(i)}]\!]$\\
       $\lss \boldsymbol{r} \rss \gets (\lss r^{((i-1)k+1)}\rss, \ldots, \lss r^{(ik)}\rss)$\\
       $\lss \boldsymbol{\alpha} \rss \leftarrow 
       \ptrunc(\lss \boldsymbol{r} \rss * ((\lss max \rss - \lss min \rss) \cdot [\![FS^{(i)}]\!])) \allowbreak + \lss min \rss \cdot [\![FS^{(i)}]\!] $
       
        \For{$\ell \gets 1$ \KwTo $k$}{
            \For{$p\gets 1$ \KwTo $n$}{
                $[\![D[p,\ell,1]]\!]_2\gets \pgeq([\![S_{\mathsf{FS}}[p,\ell]]\!],[\![\boldsymbol{\alpha}[\ell]]\!])$\\
                $[\![D[p,\ell,0]]\!]_2\gets 1 - [\![D[p,\ell,1]]\!]_2$
            }
        }
        Let $[\![t_i]\!]$ be the decision tree of depth $d$ trained using $\pSIDT$ with the data set $[\![D]\!]_2$.
    }
    \KwRet{$[\![XT]\!]=[\![t_{1}]\!], \ldots, [\![t_{m}]\!]$.}
    
    \caption{Protocol() $\pXT$ for Securely Training an Extra-Trees Classifier. 
    \label{protxt}}
\end{procedure}

Our Protocol $\pXT$ for securely training an XT classifier is given in Protocol \ref{protxt}. As with the RF approach from Section \ref{SEC:RF}, at the start of the secure XT training protocol, Alice and Bob have secret shares of the training data set $S$. 
%The class values are encoded in a simple OHE fashion, with one column for each distinct class value. An entry of one in a row represents a transaction's membership of the column's associated class. 
At the end of the protocol, they have secret shares of an XT classifier $XT=t_{1}, \ldots, t_{m}$. %trained on $S$. % Redundant?
Protocol $\ref{protxt}$ uses several building blocks that have already been introduced and explained before. 

In the offline phase, the TI generates and secret shares feature selection matrices $FS^{(1)}$, $FS^{(2)}$, \ldots, $FS^{(m)}$. The feature selection in Line 4 of Protocol \ref{protxt} is based upon the secure matrix multiplication protocol $\pmmul$, in the same way as in Section \ref{SEC:RF}. Note that here the feature selection matrices are of size $f \times k$ (with 1's representing the selections), as the features are not represented using one-hot-encodings at this point. The TI also generates the equivalent of the $r$ values from Line 5 in Algorithm \ref{FIGEXTRATREESALG} that are used for random selection of cut-off points. Note that, while in Algorithm \ref{FIGEXTRATREESALG}, each $r$ is a real number between 0 and 1, in Line 1 of Protocol \ref{protxt}, the randomly chosen values are integers between 1 and $2^a$, where $a$ is the number of fractional bits in the fixed-point representation that we use throughout this paper (see Section \ref{SEC:PRELIM}).

In Line 2 in Protocol \ref{protxt}, the parties use Protocol \ref{prot:minmax} ($\pminmax$) to compute the minimum and maximum value of each feature. In Line 5, we select a subset of the random ratios of size $k$ which will be used to calculate the vector $\lss \boldsymbol{\alpha} \rss$. This vector is calculated on Line 6, and will retain secret shared values whose sum is a random value between the minimum and maximum associated to all features that are dictated by the feature selection matrix $\lss FS^{(i)} \rss$. Note that on Line 6, $\boldsymbol{\alpha}$, $\boldsymbol{r}$ are $1 \times k$ vectors,
$max$ and $min$ are $1 \times f$ vectors, while $FS^{(i)}$ is a $f \times k$ matrix.

% As in Protocol \ref{PROT:RF}, part of the efficiency of Protocol \ref{protxt} stems from the fact that the original $n \times f$ data matrix $S$ is reduced \olive{wrong word for it, it may be expanded. Though even if it is expended, we will likely generate less splits than we would if we had to disc. per node. Sam, any thoughts?} to data matrices of size $n \times k$, and the fact that the trees can be trained in parallel to each other. The latter allows to batch operations, resulting in a communication complexity that is only dependent on the max depth of each tree, and the number of features we consider at each split.

The loop on Line 7-12 creates an one-hot-encoding of the binarized version of the data set. Each continuous feature value $S_{\rm FS}[p, \ell]$ is compared with the threshold $\alpha[\ell]$ using the secure comparison protocol $\pgeq$. The result, which is the binarized version of the data set encoded as OHE, is stored in the matrix $[\![D]\!]_2$, which is secret shared among the parties. Finally, the parties execute the secure ID3 decision tree training protocol $\pSIDT$ to jointly train a decision tree over $[\![D]\!]_2$. Similarly to Section \ref{SEC:RF}, our version of $\pSIDT$ grows each tree to the same specified depth $d$, to hide the true structure of the tree.

The two main differences between $\pXT$ and $\pRF$ are: (1) that the $\pXT$ uses (a modified version of) \textit{all} training examples for each tree, rather than using bootstrap replicas generated with bagging, and (2) that $\pXT$ splits the range of feature values by randomly selecting a single threshold. This reduces the computational complexity with respect to $\pRF$, as $\pRF$ trains $p$-nary trees while $\pXT$ only trains binary trees.

\section{Results}\label{sec:results}

\begin{table*}
\centering
\begin{tabular}{|l|r|r|c|r|r|r|}
Data set & \#instances & \#features & Model & Sklearn Acc & Secure Acc & Secure Time\\
\hline
BC & 569 & 30 & DT & 93.1\% & 90.2\% & 5.3 sec\\
   &     &    & RF & 93.7\% & 93.2\% & 18.5 sec\\  
   &     &    & XT & 96.3\% & 96.5\% & 35.2 sec\\
\hline
ECG & 14,552 & 140 & DT & 100.0\% & 100.0\% & 5.7 sec\\
    &        &     & RF & 100.0\% &  100.0\% & 79.6 sec\\
    &        &     & XT & 100.0\% & 100.0\%  & 43.6 sec\\
\hline
BACK & 310 & 12 & DT & 79.0\% & 70.0\% & 3.1 sec\\
     &     &    & RF & 82.3\% & 68.0\% & 9.8 sec\\
     &     &    & XT & 83.4\% & 81.3\% & 39.2 sec\\
\hline
IV-GSE & 225 & 12,634 & DT & 64.9\% & 59.6\% & 91.1 sec \\
       &     &        & RF & 63.6\% & 62.3\% & 12.6 sec\\
       &     &        & XT & 63.4\% & 63.5\% & 12.6 sec\\
\end{tabular}
\caption{Accuracy and runtime results for tree based models. All results are obtained with 5-fold cross-validation.}
\label{tab:mainresults}
\end{table*}

We implemented the proposed protocols in Rust\footnote{A link to the bitbucket repository is omitted to respect the double-blind review process. It will be added in the final version of the paper.} 
and performed accuracy and runtime experiments on 4 different data sets, shown in Table \ref{tab:mainresults}, namely 
the Breast Cancer\footnote{https://www.kaggle.com/uciml/breast-cancer-wisconsin-data} data set (BC), 
the ECG Heartbeat\footnote{https://www.kaggle.com/shayanfazeli/heartbeat} data set (ECG),
the Lower Back Pain Symptoms\footnote{https://www.kaggle.com/sammy123/lower-back-pain-symptoms-data set} data set (BACK) and the Track IV-GSE 2034 data set (IV-GSE) from the iDASH 2019 competition on secure genome analysis.\footnote{http://www.humangenomeprivacy.org/}
% two data sets that were made available as part of the iDASH 2019 competition on secure genome analysis\footnote{http://www.humangenomeprivacy.org/}, namely the iDASH'19 Track IV-GSE 2034 data set (IV-GSE) and the iDASH'19 Track IV-BC TCGA data set (IV-BC). 
All data sets are for binary classification problems. As shown in Table \ref{tab:mainresults}, the data sets vary in the number of continuous valued input features, as well as in the number of instances. The original ECG data set had several columns that mostly contained a value of 0, and were unhelpful to train on. We removed every feature that contained 80\% or more values of zero, reducing the ECG data set from 188 features to 140. All results in Table \ref{tab:mainresults} are obtained with 5-fold cross-validation, i.e.~the accuracies and runtimes are all averages obtained over 5 folds.

%\noindent
%\textbf{Accuracy results}\\
%\noindent
%\textbf{Accuracy Results.}
\subsection{Accuracy Results}
The ``Sklearn Acc'' column in Table \ref{tab:mainresults} contains accuracy results obtained by training tree based models over each data set in the clear, with the well known Scikit-learn library \cite{scikit-learn}. This library has state-of-the-art implementations of non privacy-preserving versions of the ML algorithms considered in Table \ref{tab:mainresults}. We used grid search to find hyperparameter values that yield good accuracy results, in line with those reported in the literature for each of the data sets; see Table \ref{tab:optimized} and below for more details. As can be observed in the Sklearn Acc column, for some data sets (e.g.~\textbf{IV-GSE}) it is substantially harder to obtain very high accuracy than for others (e.g.~\textbf{ECG}).

 The ``Secure Acc'' column in Table \ref{tab:mainresults} contains accuracy results obtained with our MPC protocols for training tree based models when the data sets are secret shared among two computing parties. The DT and RF results were obtained by running $\pdisc$ followed by respectively $\pSIDT$ and $\pRF$. The XT results were obtained by running $\pXT$. For the ring  $\mathbb{Z}_{2^\lambda}$ we used $\lambda=64$, using $a=10$ bits for the fractional part and $b=22$ bits for the integer part in the fixed point representation (see Section \ref{SEC:PRELIM}).
Computations on fixed point numbers can cause accuracy loss, particularly when products of small numbers are computed in succession. The effect is equivalent to rounding all intermediate results of a long product to some number of decimal places instead of rounding the final result. We contend with this issue by scaling the data set by a factor of 1000 before converting it to fixed point which avoids small values without requiring prior knowledge of the data. We note that scaling values in this manner does not have any effect on trees learned by ID3 in the clear.

During classification, to tally votes amongst the trees in the RF and XT classifiers trained with our secure protocols, we apply the same soft voting mechanism as what is used in Sklearn (alternative methods could be used as well, if desired). To this end, as explained in Section \ref{SEC:PRELIM}, for each classifying node, our decision tree learning protocol returns secret shared frequencies of each of the class labels in the subset of training instances that have reached that node.
In soft voting each tree returns a probability distribution of the class labels as its vote.
%Soft voting requires each tree to calculate probabilities of class labels, and return them as their vote. 
So for example, if the active classifying node in a tree has 70 positive examples, and 30 negative examples, then it would return a vote for 70\% positive, and 30\% negative. %Similarly, if a path in a tree had 7 positive examples, and 3 negative examples, it would also return a vote for 70\% positive, and 30\% negative. 
These proportions are then added up amongst all trees, and the class label with the most votes wins. 
%This was chosen because Sklearn also uses a soft voting mechanism, but in principle, we could use alternative methods if desired.

%was multiplied by 1,000 to reduce the risk of lossiness affecting our accuracy.

As we explained in prior sections, we made a variety of adaptations to the original tree model training algorithms (as implemented in Sklearn) to create MPC-friendly versions. The main high level distinction is that our protocols for DT, RF, and XT all rely on a round of discretization that creates a data set for each tree, while in Sklearn's implementations such discretization is performed for each node. In our protocols for DT and RF, our static discretization step is very explicit, as the computing parties first run $\pdisc$ to discretize the data with equal-width binning, and subsequently train a DT or a RF on the 
discretized data with $\pSIDT$ or $\pRF$. In our protocol $\pXT$ for XT, discretization is performed once for each tree, by randomly choosing feature thresholds that remain fixed for the entire tree, while in Sklearn such thresholds are randomly chosen per node. This explains most of the differences between the ``Sklearn Acc'' and ``Secure Acc'' column results in Table \ref{tab:mainresults}, along with the fact that the implementations in Sklearn are refined with some bells and whistles that we did not include in our secure version. For example, the implementation of tree learning in Sklearn has an additional stopping criterion that stops growing a tree branch if the feature values are constant across the training instances in that branch.    
%There are some notable high level differences between their algorithms and our protocols. For example, SKlearn 
%has a stopping condition where if a nodes subset of data has constant feature values, it will classify, whereas our algorithm will not classify. 
Not including this stopping criterion in our protocol was a deliberate choice to reduce communication and memory usage, but it also impacts accuracy. Despite these differences, Table \ref{tab:mainresults} convincingly shows that our protocols are competitive with the in the clear algorithms in Sklearn in terms of accuracy. 

A notable exception is the \textbf{BACK} data set, on which $\pSIDT$ and $\pRF$ clearly under-perform in terms of accuracy.
%The accuracies of our secure protocols remain competitive across with the exception of an obvious outlier, being RF on the \textbf{BACK} data set. 
The cause for this degradation in accuracy is that in $\pSIDT$ and $\pRF$ the data is first discretized with equal-width binning, and then a DT and RF are trained on the discretized data, respectively. In Sklearn on the other hand, the models are trained directly on the original, undiscretized data, with binning  performed dynamically by looking for an optimal split point in each node.
%In Sklearn, RF performs dynamic binning, whereas our protocol does binning with equal width, which severely degrades our accuracy on this data set. 
To verify this hypothesis we manually discretized the \textbf{BACK} data set with equal-width binning, and re-ran the tests with Sklearn, which gave us the same level of accuracy as obtained with $\pRF$, and even lower accuracy than $\pSIDT$. It's clear that the \textbf{BACK} data set benefits greatly from dynamic discretization. As explained in Section \ref{SEC:PROT}, \pXT offers a way to compensate for this lack of dynamic discretization by sampling the feature space more extensively and with replacement. %which is likely why \pXT performed well on it. 
The accuracy of \pXT on BACK is nevertheless still somewhat lower than the algorithms in the clear, but this is likely, as mentioned above, because the implementation of tree learning in Sklearn stops growing a branch when all feature values are constant. This stopping condition is likely to be satisfied for small data sets, and deep tree depths, and our tests on \textbf{BACK} satisfies both of these conditions (see Table \ref{tab:optimized}). On \textbf{BC} and \textbf{IV-GSE}, $\pXT$ and $\pRF$ were able to classify within $\pm 1.3\%$ of Sklearn, while $\pSIDT$ under-performed. 
% This is likely because Sklearn's implementation contains several optimizations, whereas $\pSIDT$ is a far more simple implementation of a DT. 
Lastly, on \textbf{ECG}, all protocols were able to learn the decision boundaries to perfectly separate
the positive from the negative instances.

\begin{table*}
        \centering
    \begin{tabular}{|l|r|r|r|r|r|r||r|r|r|r|r|r|r|}
    \multicolumn{7}{|c||}{Sklearn Results} & 
    \multicolumn{7}{c|}{Secure Protocol Results}\\\hline
    \hline
    Data & &  sel.feat. & trees & depth & $\epsilon$ & Acc & bins & sel.feat. & sel.inst. & trees & depth &  Time & Acc \\
    \hline
    \multirow{3}{*}\textbf{BC} & DT & -- & 1 & 4 & 5\% & 93.1\% & 5 & --& -- & 1 & 4 &  5.3 s & 90.2\% \\
    
inst: \phantom{00,}569    & RF & 17 & 70 & 4 & 5\% &  93.7\% & 6 & 30 & 200 & 100 & 3 &  18.5 s & 93.2\% \\
    
feat: \phantom{00,0}30    & XT & 19 & 90 & 5 & 5\% &  96.3\% &  -- & 128 & -- & 50 & 5 &  35.2 s & 96.5\% \\
    \hline
    
    \multirow{3}{*}\textbf{ECG}
    & DT  & -- & 1 & 1 & 5\% &  100.0\% & 2 & -- & -- & 1 & 1 &  5.7 s & 100.0\%\\
    
inst: 14,552    & RF & 120 & 20 & 1 & 5\% &  100.0\%  & 2 & 120 & 100 & 20 & 1 &  79.6 s & 100.0\% \\
    
feat: \phantom{00,}140    & XT & 120 & 20 & 1 & 5\% &  100.0\% & -- & 256 & -- & 20 & 1 & 43.6 s & 100.0\% \\
    \hline
    
    \multirow{3}{*}\textbf{BACK}
    & DT  & -- & 1 & 4 & 1\% &  79.0\% & 5 & -- & -- & 1 & 4 &  3.1 s & 70.0\% \\
    
inst: \phantom{00,}310    & RF & 8 & 90 & 5 & 1\% & 82.3\% & 5 & 10 & 30 & 100 & 5 &  9.8 s & 68.0\% \\
    
feat: \phantom{00,0}12    & XT & 10 & 120 & 6 & 1\% &  83.4\% & -- & 64 & -- & 50 & 6 &  39.2 s & 81.3\% \\
    \hline

    \multirow{3}{*}\textbf {IV-GSE}
    & DT  & -- & 1 & 1 & 5\% &  64.9\% & 5 & -- & -- & 1 & 1 &  91.1 s & 59.6\% \\
    
inst: \phantom{00,}225    & RF & 1000 & 20  & 1 & 1\% &  63.6\% & 2 & 128 & 20 & 20 & 1 &  12.6 s & 62.3\% \\
    
feat: 12,634    & XT & 112 & 10 & 3 & 1\% & 63.4\% & -- & 128 & -- & 90 & 5 &  12.6 s & 63.5\% \\
    \hline
    \end{tabular}
    \caption{Accuracy and runtime results (in seconds) for tree based models,  along with the hyperparameter values that yielded these results. We used the exact same $\epsilon$ values for the secure protocols as for the Sklearn results. All results are obtained with 5-fold cross-validation,}
    \label{tab:optimized}
\end{table*}

%\noindent
%\textbf{Runtime results}\\

%\noindent
%\textbf{Runtime Results.}
\subsection{Runtime Results}
The experiments to obtain the accuracy (Secure Acc) and runtime (Secure Time) results were run on Microsoft Azure Lv48s\_vs machines with 48 vCPUs, 384.0 GiB Memory. Each of the parties ran on a separate machine (connected via Gigabit Ethernet network) 
which means that the results in Table \ref{tab:mainresults} cover communication time in addition to computation time. All reported runtimes are for the online phases of the protocols. As can be seen, our protocols are very fast, even on data sets with 1000s of instances or features.

We repeat the accuracy and runtime results from Table \ref{tab:mainresults} in Table \ref{tab:optimized}, along with our choices for the hyperparameter values. We obtained these hyperparameter values by performing grid search while aiming for state-of-the-art accuracies on the respective data sets. The attentive reader will notice that the hyperparameter values differ between the algorithms as implemented in Sklearn, and our MPC-based protocols. As we explain in more detail below, the incentive to choose hyperparameter values differently for the MPC-based protocols goes hand in hand with the adaptations made to make the original ML algorithms more MPC-friendly. 

%All accuracy and runtime results were obtained with 5-fold cross-validation, where each fold was ran 5 times, totalling in 25 runs each. 

For the Sklearn results in Table \ref{tab:optimized}, the hyperparameter \textit{sel.~feat.} denotes how many random features were selected for each tree in the RF and XT classifiers (the original number of features for each data set is recalled in the first column of Table \ref{tab:optimized}). \textit{Trees} and \textit{depth} denote how many trees were trained in each ensemble, and to what depth each tree was to be trained to. Finally, $\epsilon$ denotes what fraction of training instances needed to remain in a node for that node to branch out further. For example, if we have a data set of 1000 instances, and $\epsilon = 5\%$, then if the number of training instances that have reached a node during tree construction is $50$ instances or less, the growing stops in that branch and the node becomes a classifying node. We used the same $\epsilon$ values for the secure protocols as for Sklearn (not repeated in the table for conciseness). 

There are two additional columns of hyperparameters in Table \ref{tab:optimized} for the secure protocols, namely \textit{bins}, which is the number of buckets of equal width that $\pdisc$ generates for DT and RF, and \textit{sel.~inst.}, which denotes how many random instances where selected, per tree and with replacement, from each discretized set by $\pRF$. For the Sklearn implementation, we followed the convention of choosing \textit{sel.~inst.} equal to the total number of instances (which is recalled for each data set in the first column of Table \ref{tab:optimized}) while for $\pRF$ we systematically chose a smaller value for \textit{sel.~inst.} for efficiency reasons, as explained in Section \ref{SEC:RF}.

%Table \ref{tab:optimized} shows our runtimes, and how our accuracies fare against Sklearn's accuracies given different hyperparameters. The hyperparameter \textit{bins} denote how many buckets of equal width our secure RF and DT algorithm generate. The hyperparameters \textit{sel. feat.} and \textit{sel. inst.} denote how many random features and random instances where selected respectively for each discretized set.  Finally, $\epsilon$ tells us what proportion of the original data set has to reach for a node to satisfy an early exit condition. So, for example, if we have a data set of 1000 instances, and $\epsilon = 5\%$, then if a node's subset of the data is $50$ instances or less the node should classify. 

% The hyperparameters between Sklearn's results and our own differ to varying degrees. The main difference between the hyperparemeters of our secure DT and Sklearn's DT is that we get to choose how we bin data, while Sklearn performs this dynamically. Thus, our hyperparameters are very similar. In contrast, the implementations of our secure XT and RF, and Sklearns vary in more significant ways. Our RF algorithm can vary in the number of samples it bootstraps, and our XT algorithm collects a pool of attributes without replacement. Given these changes, among other internal differences of how our algorithms process data, our hyperparameters vary. ~~~ Possible discussion about why our hyperparameters differ from sklearns. Not sure how valuable it is

Our runtimes for DT are far shorter than those of XT and RF on the data sets with a small number of features, namely \textbf{BC}, \textbf{ECG}, and \textbf{BACK}, and much longer on data set \textbf{IV-GSE}, which has +12K features. This is entirely with expectations, as $\pRF$ and $\pXT$ have a built-in mechanism for random feature selection, which can greatly reduce the number of features that need to be considered per tree without harming the accuracy, while $\pSIDT$ has to consider all features. 

On the flip-side, this random feature selection in $\pRF$ and $\pXT$ does not come for free, and neither does the random instance selection in $\pRF$. As explained in Section \ref{SEC:RF} and \ref{SEC:PROT}, such oblivious extraction of subsets of the data is realized through secure matrix multiplication.
%For \textbf{BC}, \textbf{ECG}, and \textbf{BACK} our runtimes for DT are far shorter than that of XT and RF. This is because how fast the pre-processing phase is. We don't have to extract subsets of data, and as such, we get to sidestep costly matrix multiplication operations. Thus, the runtime of our secure DT protocol is mostly determined by the training phase. 
%This also explains why DT took longer on the \textbf{IV-GSE} data set than RF and XT. In our tests the amount of selected features for RF and XT were far less than the total amount available, alleviating stress on the training phase. The DT algorithm instead trains a tree on all features, adding considerably to the runtime. Note that in practice, single DTs need to be grown fairly deep to properly learn decision boundaries of the data. In our tests, DT had at most a depth of 4, making it relatively shallow. Given the exponential growth of local operations, bandwidth, and memory in \pSIDT, a deep DT may be infeasible. 
%For the data sets \textbf{BC}, \textbf{ECG}, and \textbf{BACK}, XT and RF took relatively long to perform. The XT algorithm requires us to obliviously select features using matrix multiplication, whereas RF requires us to obliviously select features and samples, which requires two separate matrix multiplications. 
This matrix multiplication only requires a single round of communication, but the local complexity is quite high. Further optimization could be done by offloading the work onto a GPU.

%Ideally, we would offload this work onto a GPU, but we did not have access to a GPU in our tests, making \pmmul an expensive protocol. Regardless, the performance of all of our protocols in our tests are very fast, never taking more than two minutes. With our competitive accuracies and quick performance, our work is fast enough to be used in practice.

Recall from Section \ref{SEC:PROT} that to compensate for the static discretization per tree, in $\pXT$ the parties jointly
%give $\pXT$ a more diverse range of choices, we 
select features \textit{with replacement}, and choose a different split for each. This explains why \textit{sel.~feat.} for XT in the secure protocol results columns in Table \ref{tab:optimized} can  exceed the total amount of features in the data set. This design choice benefits \pXT greatly, making it fast and the most competitive with Sklearn's implementation in terms of accuracy. Given our choice to select a pool of random features per tree as opposed to per node, we also save valuable computation time. Take our \textbf{BC} results for XT as an example. In Sklearn's implementation, each node must generate 19 unique features. If each of their trees are fully grown, as ours must be to protect from side channel attacks, they would have to generate $(2^5 - 1) \cdot 19 = 589$ total features and cut-off points. In contrast, we generate a pool of 128 features up-front, saving us an immense amount of time, without harming accuracy.

\begin{table}
\centering
\begin{tabular}{ |c|c|c|c|c|} 
 \hline
 \# of elements & $10^6$ & $10^7$ & $10^8$ & $10^9$  \\ 
 \hline
 \ptoq & 0.1 sec & 0.4 sec & 3.7 sec & 58.1 sec  \\ 
 \pgeq & 0.4 sec & 2.0 sec & 17.0 sec & 152.2 sec   \\ 
 \pminmax & 1.9 sec & 7.1 sec & 50.1 sec & 572.7 sec  \\ 
 \hline
\end{tabular}
\caption{Runtimes of protocol executions on varying input sizes. \pgeq compares two vectors of size \# elements while \ptoq and \pminmax perform operations on one vector of size \# elements. }
\label{tab:vecRuntimes}
\end{table}

\begin{table}
\centering
\begin{tabular}{ |c||c|c|c|} 
 \hline
 &
 \multicolumn{3}{c}{Instances} \\
 \hline
  Bins & $10^3$ & $10^4$ & $10^5$  \\ 
 \hline
 2 & 2.3 sec & 13.6 sec & 106.5 sec   \\ 
 3 & 2.8 sec & 19.9 sec & 139.6 sec   \\ 
 5 & 3.3 sec & 24.1 sec & 199.0 sec   \\ 
 8 & 9.2 sec & 28.8 sec & 374.1 sec   \\ 
 \hline
\end{tabular}
\caption{Runtime of \pdisc for a varying number of instances and bins, and a fixed amount of 1,000 features}
\label{tab:discRuntimes}
\end{table}

Tables \ref{tab:vecRuntimes} and \ref{tab:discRuntimes} show the time it takes to perform the subprotocols for multiple input lengths. Table \ref{tab:vecRuntimes} shows the runtimes of \ptoq, \pgeq, and \pminmax up to $10^9$ elements, where \ptoq and \pminmax performed their operations on a single vector of that length, and \pgeq performed operations between two vectors of that length. We stop at $10^9$ because that is the point our virtual machines run out of RAM. Table \ref{tab:discRuntimes} shows the runtimes for \pdisc for a varying number of bins and instances, where the amount of features is fixed at 1,000. These runtimes are quite fast compared to previous iterations and are notably sublinear over certain ranges. This is the result of optimized multithreading and socket-level parallelization.

\section{Conclusion}

% \magenta{Here we should say which method we would choose if we were to train a tree based model on data that we can not see. Say that we do not have the liberty to first bring the data together in the clear and play with all the hyperparameter choices, which method would we choose (DT, RF, XT), and how would we choose values for the hyperparameters? I think that it is a strong selling point that our training times are very fast, which means that we could train a variety of models and then evaluate them in a privacy-preserving manner to select the best one!}

In this paper we have presented several cryptographic protocols that enable two or more parties to train decision trees or decision tree ensemble classifiers over their joint continuous valued data, while keeping the data private, i.e.~neither of the parties has to show their data to anyone in the clear. Our results demonstrate that it is possible to efficiently train tree-based models with good accuracy while completely avoiding a full private sorting, by performing various forms of discretization in a privacy-preserving manner instead. 

%Two of our proposals work by using private discretization with equal binning, and applying the discretized data set to a modified version of the protocol proposed in \cite{de2014practical}. The other proposal modifies the extra-trees machine learning algorithm to a privacy-preserving setting, requiring us to select random cut-off points, ultimately discretizing the data, and allowing us to use the protocol the modified protocol in \cite{de2014practical} again.

The results of our protocols demonstrate competitive accuracy with their state-of-the-art, in the clear counterparts. Beyond accuracy, each one of our protocols demonstrated high efficiency in terms of runtime, and is fast enough to be used in practice. Our MPC protocol for extra-trees classifiers is the clear winner among our three secure approaches for training tree based models on continuous valued data, as it is both accurate and fast. When confronted with the need to securely train a tree-based model over data that is distributed across different parties, we therefore recommend protocol $\pXT$ as the method of choice.

An important question during such deployment is which values to choose for the hyperparameters, such as the number of trees and the number of randomly selected features per tree. Given the high efficiency of our protocols, it would be feasible to train multiple models with multiple different hyperparameters jointly amongst the parties.  Then, in conjunction with protocols for privacy-preserving inference \cite{IEEETDSC:CDHK+17,fritchman2018privacy}, we could perform a secure comparison of all of the results, obliviously select the most accurate model with the best hyperparameters, and use that to classify new, unseen instances.

%In particular, the accuracy of our secure extra-trees algorithm performs most consistently with its counterpart, only suffering a small decrease in accuracy for one of the data sets. 

%In real world applications, we don't have the luxury of combining data sets from multiple parties in the clear, and playing around with the hyper parameters of different models to achieve the best possible accuracy. Given the speed of our results, it would be feasible to train multiple models with multiple different hyperparameters jointly amongst the parties. Then, in conjunction with privacy-preserving classification protocols such as the one described in \cite{IEEETDSC:CDHK+17} (\olive{Is this the right citation?}), we could perform a secure comparison of all of the results, and obliviously select the most accurate model with the best hyperparameters, and use that to classify new, unseen instances.

The accuracy of machine learning algorithms can be highly dependent on specific properties of data sets. Thus, it is important to have a portfolio of privacy-preserving machine-learning algorithms readily available for use. The presented protocols fill an important gap in the literature, presenting the first protocols for trees and tree ensembles that can handle continuous data without relying on a full sorting of the data set.

\appendix

\section{Security Model and Proofs}\label{sec:security}

The security model considered in this work is the Universal Composability (UC) framework \cite{FOCS:Canetti01}, the gold standard for formally defining and analyzing the security of cryptographic protocols. Any protocol that is proven UC-secure, can be arbitrarily composed with other copies of itself and of other protocols (even with arbitrarily concurrent executions) while preserving security. That is an extremely useful property that allows the modular design of cryptographic protocols. UC-security is also a necessity for cryptographic protocols running in complex environments such as the Internet. This appendix only gives a short overview of the UC framework for the specific case of protocols with two participants (denoted Alice and Bob). See \cite{CDN2015} for more details. 

In the UC framework the security is analyzed by comparing a real world with an ideal world. In the real world Alice and Bob interact between themselves and with an adversary $\adv$ and an environment $\env$. The environment $\env$ captures all external activities to the protocol instance under consideration, and is responsible for giving the inputs and getting the outputs from Alice and Bob. The adversary $\adv$ can corrupt either Alice or Bob, in which case he gains the control over that participant. The network scheduling is assumed to be adversarial and thus $\adv$ is responsible for delivering the messages between Alice and Bob. In the ideal world, there is an ideal functionality $\F$ that captures the perfect specification of the desired outcome of the computation. $\F$ receives the inputs directly from Alice and Bob, performs the computations locally following the primitive specification and delivers the outputs directly to Alice and Bob. A protocol $\pi$ executed between Alice and Bob in the real world UC-realizes the ideal functionality $\F$ if for every adversary $\adv$ there exists a simulator $\s$ such that no environment $\env$ can distinguish between: (1) an execution of the protocol $\pi$ in the real world with participants Alice and Bob, and adversary $\adv$; (2) and an ideal execution with dummy parties (that only forward inputs/outputs), $\F$ and $\s$.

We design our protocols in the trusted initializer (TI) model, which is formalized by the trusted initializer functionality $\fti{}$. The TI pre-distributes correlated randomness to Alice and Bob, but neither takes part in any part of the protocol execution nor learns any inputs or outputs of Alice and Bob. \footnote{Setup assumptions are necessary to obtain UC-security \cite{C:CanFis01,STOC:CLOS02}. Some other possible setup assumption for obtaining UC-security are: a common reference string \cite{C:CanFis01,STOC:CLOS02,C:PeiVaiWat08}, the random oracle model \cite{TCC:HofMul04,EPRINT:BDDMN17b}, the availability of a public-key infrastructure \cite{FOCS:BCNP04}, the existence of noisy channels between the parties \cite{SBSEG:DMN08,JIT:DGMN13}, and the availability of tamper-proof hardware \cite{EC:Katz07,ICITS:DowMulNil15}.} A TI was used before to enable highly practical solutions both in the context of PPML, e.g., \cite{AISec:CDNN15,david2015efficient,fritchman2018privacy,IEEETDSC:CDHK+17,IEEENSRE:ADMW+19,NeurIPS2019,idash} as well as in other important applications, e.g., \cite{r99,dowsley2010two,IEICE:DMOHIN11,ishai2013power,IJIS:TNDMIHO15,IEEEIFS:DDGM+16}. 

\begin{functionality}{Functionality $\fti{}$}
$\fti{}$ is parametrized by algorithm $\mathcal{D}$ that samples correlated randomness. Upon initialization, run $(D_A, D_B) \getsr \mathcal{D}$, and deliver $D_A$ to Alice and $D_B$ to Bob.
\end{functionality}

\textbf{Simplifications:} The messages of ideal functionalities are formally public delayed outputs, meaning that $\s$ is first asked whether they should be delivered or not (this is due to the modeling that the adversary controls the network scheduling). This detail as well as the session identifications are omitted from the description of our functionalities for the sake of readability. 

Our protocols are information-theoretically secure and the simulation strategy is quite simple: all the messages look uniformly random from the recipient's point of view, except for the messages that open a secret shared value to a party, but these ones can be easily simulated using the output of the respective functionalities. A simulator $\s$, having the leverage of being able to simulate the ideal functionalities that capture the trusted initializer functionality $\fti{}$ and the ideal functionalities that are UC-realized by the sub-protocols, can easily extract the necessary values and perform a perfect simulation of a real protocol execution; therefore making the real and ideal worlds indistinguishable for any environment $\env$. The simulation strategy will be described briefly in our proofs.

The protocol for secure matrix multiplication $\pmmul$ UC-realizes the distributed matrix multiplication functionality $\fmmul$ \cite{Dowsley16,IEEETDSC:CDHK+17} in the trusted initializer model.

\begin{functionality}{Functionality $\fmmul{}$}
$\fmmul$ is parametrized by the size $q$ of the ring $\mathbb{Z}_q$ and the dimensions $(i, j)$ and $(j, k)$ of the matrices.
\vspace{1mm}

\textbf{Input:} Upon receiving a message from Alice/Bob with its shares of $\shareq{X}$ and $\shareq{Y}$, verify if the share of $X$ is in $\Zqm{i}{j}$ and the share of $Y$ is in $\Zqm{j}{k}$.
If it is not, abort. Otherwise, record the shares, ignore any subsequent message from that party and
inform the other party about the receipt.
\vspace{1mm}

\textbf{Output:} Upon receipt of the shares from both parties, reconstruct $X$ and $Y$ from 
the shares, compute $Z=X Y$ and create a secret sharing $\shareq{Z}$ to distribute to Alice and Bob: a corrupt party fixes its share of the output to any chosen matrix and the shares of the uncorrupted parties are then created by picking uniformly random values subject to the correctness constraint.
\end{functionality}

The protocol \ptoq for converting a secret bit from a $\mathbb{Z}_2$ sharing to a $\mathbb{Z}_q$ sharing
UC-realizes the share conversion functionality $\fconv$ \cite{NeurIPS2019}.

\begin{functionality}{Functionality $\fconv$}
$\fconv$ is parametrized by the size of the field $q$.
\vspace{1mm}

\textbf{Input:} Upon receiving a message from Alice/Bob with her/his share of $\sharetwo{x}$, record the share, ignore any subsequent messages from that party and inform the other party about the receipt.
\vspace{1mm}

\textbf{Output:} Upon receipt of the inputs from both parties, reconstruct $x$, then create and distribute to Alice and Bob the secret sharing $[\![x]\!]_{q}$. Before the deliver of the output shares, a corrupt party fix its share of the output to any constant value. In both cases the uncorrupted parties' shares are then created by picking uniformly random values subject to the correctness constraint.
\end{functionality}

The bit extraction protocol \pbtx is a straightforward simplification of the bit decomposition protocol \pdecompopt from \cite{idash} and UC-realizes the bit extraction functionality $\fbtx$.

\begin{functionality}{Functionality $\fbtx$}
$\fbtx$ is parametrized by the bit-length $\lambda$ of the secret shared input value $x$ and extracts the $\alpha$-th bit from the secret shared value.
\vspace{1mm}

\textbf{Input:} Upon receiving a message from Alice or Bob with its share of $[\![x]\!]$, record the share, ignore any subsequent messages from that party and inform the other party about the receipt.
\vspace{1mm}

\textbf{Output:} Upon receipt of the inputs from both parties, reconstruct the value $x=x_\lambda \cdots x_1$ from the shares, and distribute a new secret sharing $\sharetwo{x_\alpha}$ of the bit $x_\alpha$. Before the output delivery, the corrupt party fixes its shares of the output to any desired value. The shares of the uncorrupted parties are then chosen uniformly at random subject to the correctness constraints.
\end{functionality}

The secure comparison \pgeq protocol is trivially correct. 
The simulator $\s$ internally simulates an execution of Protocol \pgeq for the adversary $\adv$ that controls the corrupted party. Using the fact that he is the one simulating the trusted initializer in this execution, $\s$ can extract the shares of the inputs $x$ and $y$ that belong to the corrupted party and forward them to $\fgeq$. $\s$ can then fix in $\fgeq$ the corrupted party's share of the output to the value that matches what $\adv$ gets in the simulated execution of \pgeq. The simulation is perfect, and therefore the environment $\env$ cannot distinguish the real and ideal worlds, and \pgeq UC-realizes \fgeq.

\begin{functionality}{Functionality $\fgeq$}
$\fgeq$ runs with Alice and Bob and is parametrized by the bit-length $\lambda$ of the values $x$, $y$ to be compared. $x$ and $y$ are guaranteed to be such that $|x-y|< 2^{\lambda-1}$.
\vspace{1mm}

\textbf{Input:} Upon receiving a message from Alice or Bob with its share of $[\![x]\!]$ and $[\![y]\!]$, record the shares, ignore any subsequent messages from that party and inform the other party about the receipt.
\vspace{1mm}

\textbf{Output:} Upon receipt of the inputs from both parties, reconstruct the values $x$ and $y$. If $x \geq y$, distribute a new secret sharing $\sharetwo{1}$; otherwise a new secret sharing $\sharetwo{0}$. Before the output deliver, the corrupt party fix its shares of the output to any desired value. The uncorrupted parties' shares are created by picking uniformly random values subject to the correctness constraints.
\end{functionality}

The secure equality \peq protocol is also trivially correct and the simulator for \feq uses a similar strategy as the one for \fgeq to get a perfect simulation. Thus \peq UC-realizes \feq.

\begin{functionality}{Functionality $\feq$}
$\feq$ runs with Alice and Bob and is parametrized by the bit-length $\lambda$ of the values $x$, $y$ to be compared. $x$ and $y$ are guaranteed to be such that $|x-y|< 2^{\lambda-1}$.
\vspace{1mm}

\textbf{Input:} Upon receiving a message from Alice or Bob with its share of $[\![x]\!]$ and $[\![y]\!]$, record the shares, ignore any subsequent messages from that party and inform the other party about the receipt.
\vspace{1mm}

\textbf{Output:} Upon receipt of the inputs from both parties, reconstruct the values $x$ and $y$. If $x = y$, distribute a new secret sharing $\sharetwo{1}$; otherwise a new secret sharing $\sharetwo{0}$. Before the output deliver, the corrupt party fix its shares of the output to any desired value. The uncorrupted parties' shares are created by picking uniformly random values subject to the correctness constraints.
\end{functionality}

It is trivial to verify that the protocol \pminmax correctly outputs secret sharings corresponding to the minimum and maximum values of the input vector. The simulator $\s$ uses the fact that he is responsible for simulating \fgeq in the internal simulated execution of \pminmax with \adv in order to extract the corrupted party's shares of $d_1, \ldots, d_n$. $\s$ can then forward those input shares to \fminmax and adjust the shares of the outputs $d_{min}$ and $d_{max}$ in the functionality to match what \adv sees in the simulated execution of \pminmax. The simulation is perfect and \pminmax UC-realizes \fminmax.

\begin{functionality}{Functionality $\fminmax$}
$\fminmax$ is parametrized by the size $n$ of the vector.
\vspace{1mm}

\textbf{Input:} Upon receiving a message from Alice or Bob with its shares of $[\![d_1]\!], \ldots, [\![d_n]\!]$, record the shares, ignore any subsequent messages from that party and inform the other party about the receipt.
\vspace{1mm}

\textbf{Output:} Upon receipt of the inputs from both parties, reconstruct the values $d_1, \ldots, d_n$. Compute the minimum $d_{min}$ and maximum $d_{max}$ values contained in the vector. Distribute new secret sharings $[\![d_{min}]\!]$ and $[\![d_{max}]\!]$. Before the output deliver, the corrupt party fix its shares of the output to any desired value. The uncorrupted parties' shares are created by picking uniformly random values subject to the correctness constraints.
\end{functionality}

It is easy to verify that the secure discretization protocol \pdisc correctly computes the bins thresholds and the one-hot encodings of the bin membership of each $d \in D$. The simulator $\s$ for $\fdisc$ internally runs an execution of $\pdisc$ for $\adv$ and uses the fact that he simulates $\fminmax$ in order to extract the corrupted party's shares of the values $d_1, \ldots, d_n$ that he needs to forward to \fdisc. It can then trivially adjust the corrupted party's output shares in \fdisc to matches the ones in the simulated execution of \pdisc. This is a perfect simulation strategy and \pdisc UC-realizes \fdisc.

\begin{functionality}{Functionality $\fdisc$}
$\fdisc$ is parametrized by the number of elements $n$ in the vector $D$ and the number of bins $p$.
\vspace{1mm}

\textbf{Input:} Upon receiving a message from Alice or Bob with its shares of $[\![D]\!]$, record the shares, ignore any subsequent messages from that party and inform the other party about the receipt.
\vspace{1mm}

\textbf{Output:} Upon receipt of the inputs from both parties, reconstruct the values $d_1, \ldots, d_n$. Compute the minimum $d_{min}$ and maximum $d_{max}$ values contained in the vector, and the thresholds of the $p$ equal-width bins. Map the values $d \in D$ to the respective bins and create $D'$ that contains one-hot-encodings of the bin membership of each $d \in D$. 
Distribute $[\![D']\!]_2$. Before the output deliver, the corrupt party fix its shares of the output to any desired value. The shares of the uncorrupted parties are then created by picking uniformly random values subject to the correctness constraints.
\end{functionality}

We also use as a building block a protocol $\pSIDT$ that UC-realizes a decision tree training functionality $\fSIDT$. In particular, we use a slightly modified version of the protocol by De Hoogh et al. \cite{de2014practical}.

\begin{functionality}{Functionality $\fSIDT$} \fSIDT runs with Alice and Bob to train a decision tree model with categorical values. It is parametrized by the number of samples $n$ and of features $f$ in the data set $S$, the number $p$ of categories per feature, and the depth $d$ of each tree. The input data is presented in a one-hot-encoding (OHE) format and the Gini index is used to select the split features.
\vspace{1mm}

\textbf{Input:} Upon receiving a message from Alice or Bob with its shares of $[\![S]\!]_2$, record the shares, ignore any subsequent messages from that party and inform the other party about the receipt.
\vspace{1mm}

\textbf{Output:} Upon receipt of the inputs from both parties, reconstruct the data set $S$ and locally train a decision tree $t$ using the same criteria and representation format as \pSIDT. Distribute $[\![t]\!]$. Before the output deliver, the corrupt party fix its shares of the output to any desired value. The shares of the uncorrupted parties are then created by picking uniformly random values subject to the correctness constraints.
\end{functionality}

When we concatenate the protocols \pdisc and \pSIDT, they do not leak any information due to the UC-security of their building blocks. This concatenated protocol UC-realizes \fdiscdt. The simulator $\s$ can easily extract the corrupted party's shares of the data set by using the fact that it simulates \fdisc in the internal simulated execution with the adversary $\adv$. $\s$ can provide this information to $\fdiscdt$ and adjust the outputs of $\fdiscdt$ to match the internal execution with \adv.

\begin{functionality}{Functionality $\fdiscdt$}
$\fdiscdt$ runs with Alice and Bob and first performs a discretization of the continuous-valued data set $S$ and then trains a decision tree on the categorical data.  It is parametrized by the number of samples $n$ and of features $f$ in the data set $S$, the number of bins $p$ per feature and the depth $d$ of each tree.
\vspace{1mm}

\textbf{Input:} Upon receiving a message from Alice or Bob with its shares of $[\![S]\!]$, record the shares, ignore any subsequent messages from that party and inform the other party about the receipt.
\vspace{1mm}

\textbf{Output:} Upon receipt of the inputs from both parties, compute locally the discretization of $S$ followed by the secret shared tree $[\![t]\!]$ using the same procedures as $\pdisc$ followed by $\pSIDT$ (the necessary correlated randomness is generated locally). Distribute $[\![t]\!]$. Before the output deliver, the corrupt party fix its shares of the output to any desired value. The shares of the uncorrupted parties are then created by picking uniformly random values subject to the correctness constraints.
\end{functionality}

The protocol \pRF for training a random forest on data that is initially discretized clearly does not leak any information due to the UC-security of its building blocks. The simulator $\s$ can easily extract the corrupted party's shares of the data set by using the fact that it simulates \fdisc in the internal simulated execution of \pRF with the adversary $\adv$. $\s$ can provide this information to $\fdiscrf$ and adjust the outputs of $\fdiscrf$ to match the internal execution of \pRF, thus \pRF UC-realizes \fdiscrf.

\begin{functionality}{Functionality $\fdiscrf$}
$\fdiscrf$ runs with Alice and Bob and first performs a discretization of the continuous-valued data set $S$ and then trains a random forest on the categorical data.  It is parametrized by the number of samples $n$ and of features $f$ in the data set $S$, the number of bins $p$ per feature, the number $k$ of features to use in each tree, the number of samples $s$ used in each tree, the number $m$ of trees in the ensemble and the trees' depth $d$.
\vspace{1mm}

\textbf{Input:} Upon receiving a message from Alice or Bob with its shares of $[\![S]\!]$, record the shares, ignore any subsequent messages from that party and inform the other party about the receipt.
\vspace{1mm}

\textbf{Output:} Upon receipt of the inputs from both parties, compute locally the discretization of $S$ followed by the secret shared trees $[\![t_1]\!],\ldots, [\![t_m]\!]$ using the same procedures as $\pRF$ (the necessary correlated randomness is generated locally). Distribute $[\![RF]\!]=[\![t_1]\!],\ldots, [\![t_m]\!]$. Before the output deliver, the corrupt party fix its shares of the output to any desired value. The uncorrupted parties' shares are created by picking uniformly random values subject to the correctness constraints.
\end{functionality}

The protocol \pXT for training an extra-trees model clearly does not leak any information due to the UC-security of its building blocks. The simulator $\s$ can easily extract the corrupted party's shares of the data set by using the fact that it simulates \fmmul in the internal simulated execution of \pXT with the adversary $\adv$. $\s$ can provide this information to $\fXT$ and adjust the outputs of $\fXT$ to match the internal execution of \pXT, therefore \pXT UC-realizes \fXT.

\begin{functionality}{Functionality $\fXT$}
$\fXT$ runs with Alice and Bob to train an extra-trees model. It is parametrized by the number of samples $n$ and of features $f$ in the data set $S$, the number $k$ of features to use in each tree, the number $m$ of trees in the ensemble and the depth $d$ of each tree.
\vspace{1mm}

\textbf{Input:} Upon receiving a message from Alice or Bob with its shares of $[\![S]\!]$, record the shares, ignore any subsequent messages from that party and inform the other party about the receipt.
\vspace{1mm}

\textbf{Output:} Upon receipt of the inputs from both parties, compute the secret shared trees $[\![t_1]\!],\ldots, [\![t_m]\!]$ locally using the same procedures as $\pXT$ (the necessary correlated randomness is generated locally). Distribute $[\![XT]\!]=[\![t_1]\!],\ldots, [\![t_m]\!]$. Before the output deliver, the corrupt party fix its shares of the output to any desired value. The uncorrupted parties' shares are chosen uniformly at random  subject to the correctness constraints.
\end{functionality}

\section{Efficient Bit Extraction}\label{sec:btx}

The protocol for secure bit extraction, \pbtx, is an of the matrix composition network approach to bit decomposition presented in \cite{idash} as \pdecompopt. Here, we summarize the notions used to derive \pdecompopt and then derive our modification \pbtx.

\subsection{Optimized Bit Decomposition Protocol Using Matrix Composition}

The decomposition of a secret-shared value $\lss x \rss_{2^\lambda} = a + b \mod 2^\lambda$ into bitwise shares in $\mathbb{Z}_2$ is modeled as an adder circuit where, beginning with the least significant bit of $a, b$, all subsequent bits are computed as a bitwise sum plus the carry-over from the previous bitwise sum. Formally, we consider two signals which depend on $a, b$: $\mathsf{Generate}$ ($g_i = a_ib_i$) generates a carry bit at the $i$-th position, and $\mathsf{Propagate}$ ($p_i = a_i \oplus b_i$) propagates a carry bit, if it exists \cite{idash}. Then, the $i$-th bit of the sum is given by $s_i = p_i \oplus c_{i-1}$ where the $i$-th \textit{carry bit} is given by $c_{i} = g_i + p_i c_{i-1}$. 

It was noted in \cite{idash} that a matrix representing this formula for the carry bit can be computed efficiently in the MPC setting with minimal dependency on the results for previous carry bits. Moreover, all carry bits can be computed in advance by composing entries of the matrix representation. Let $M_i$ denote the matrix that holds the $i$-th $\mathsf{Generate}$ and $\mathsf{Propagate}$ bits.

\[
\begin{bmatrix}
c_i \\ 1
\end{bmatrix}
=
\begin{bmatrix}
p_i & g_i \\
0 & 1 \\
\end{bmatrix}
\begin{bmatrix}
c_{i-1} \\ 1
\end{bmatrix}
=
M_i
\begin{bmatrix}
c_{i-1} \\ 1
\end{bmatrix}.
\]  

\begin{figure*}
\centering
    \includegraphics[width=0.85\textwidth]{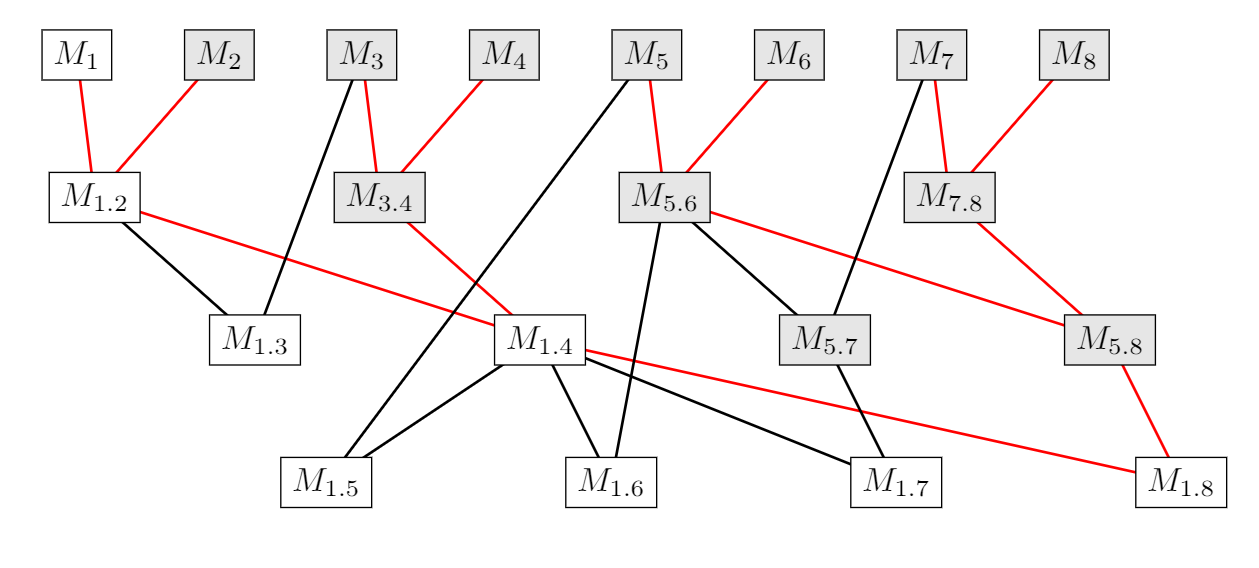}
    \caption{$\textsf{ComposeNet}_{\lambda}$ for $\lambda=9$. Computes the set of all matrix compositions $M_1, M_{1.2}, M_{1.3},\ldots,M_{1.(\lambda-1)}$  \cite{idash}. The subgraph denoted in red computes $\textsf{ComposeNet}^\mathsf{BTX}_{\alpha}$ for $\alpha=9$; the matrix composition $M_{1.(\alpha-1)}$.}
    \label{fig:matcompbtx}
\end{figure*}

Then, all carry bits $c_i$ can be derived by the product $M_i \cdot ( c_{i-1} , 1 )$, but they can be equally be computed by the composition of all matrices $M_iM_{i-1}\dots{M_1} \cdot (0, 1)$. Further, by noting that the product
\[
M_i M_{i-1} \dots M_1 
\begin{bmatrix}
0 \\ 1
\end{bmatrix}
=
\begin{bmatrix}
p'_i & g'_i \\
0 & 1 \\
\end{bmatrix}
\begin{bmatrix}
0 \\ 1
\end{bmatrix}
=
\begin{bmatrix}
g'_i \\ 1
\end{bmatrix},
\] 
the $i$-th carry bit $c_i$ is implicitly the upper right-hand entry of the composed matrix, $g'_i$. 
Given that matrix composition is left-associative, the set of all matrices needed for the final result can be computed in a log depth circuit -- hereafter $\textsf{ComposeNet}_{\lambda}$ --  by, at the $i$-th layer, computing all compositions $M_{1.j}$ that require fewer than $2^{i-1}$ composition operations. The constraint is added that each $M_{1.j}$ should be the composition of the ``largest'' matrix from the previous layer, $M_{1.2^{i-2}}$, with the remainder $M_{2^{i-2}+1.j}$. If $M_{2^{i-2}+1.j}$ doesn't exist in the network, it is added recursively \cite{idash}. For a ring element with bit length $\lambda$, the depth of $\textsf{ComposeNet}_{\lambda}$ is $\lceil \log(\lambda - 1) \rceil$. This is because the $\lambda$-th sum bit depends only on $c_{\lambda-1}$. An example of $\textsf{ComposeNet}_{\lambda}$ for $\lambda=9$ is given in Figure \ref{fig:matcompbtx}. The optimized bit decomposition protocol $\pdecompopt$ of \cite{idash} is described in Protocol \ref{prot:decompopt}.

\begin{procedure}[h]
    \SetKwInOut{Input}{Input}
    \SetKwInOut{Output}{Output}

    \Input{ $[\![x]\!]_{2^{\lambda}}  $}
    \Output{$[\![x_1]\!]_2...[\![x_{\lambda}]\!]_2$}
    
    Party $i$ regards its share $x_i$ as $p_{i,1}, \ldots,p_{i,\lambda}$ s.t. \\
    $[\![ p_j ]\!]_2, = p_{1,j} \oplus p_{2,j}$ for $j=1,\ldots, \lambda$ \\
    Party 1 creates the sharing $[\![ g_{1,j} ]\!]_2=(p_{1,j}, 0)$.\\
    Party 2 creates the sharing $[\![ g_{2,j} ]\!]_2=(0,p_{2,j})$.\\
    $[\![ g_j ]\!]_2 \leftarrow [\![ g_{1,j} ]\!]_2[\![ g_{2,j} ]\!]_2$\\
    $[\![M_j]\!]_2 \leftarrow 
        \begin{bmatrix}
            [\![p_j]\!]_2 & [\![g_j]\!]_2 \\
            0 & 1 \\
        \end{bmatrix}$ for all $j$\\
    $\{ [\![M_{1.j}]\!]_2 \; | \; 1\leq j < \lambda\} \leftarrow 
    \mathsf{ComposeNet}_{\lambda}( [\![M]\!]_2 )$ \\
    $[\![c_j]\!]_2 \leftarrow$ the upper right entry of $[\![M_{1.j}]\!]_2$ \\
    $[\![s_1]\!]_2 \leftarrow [\![p_1]\!]_2$  \\
    $[\![s_j]\!]_2 \leftarrow [\![p_j]\!]_2 \oplus [\![c_{j-1}]\!]_2$ for all $j > 1$ \\

    \KwRet{$[\![s_1]\!]_2...[\![s_{\lambda}]\!]_2$}
    \caption{Secure Optimized Bit Decomposition Protocol() $\pdecompopt$.}\label{prot:decompopt} 
\end{procedure}

\subsection{Bit Extraction Protocol}

A key aspect of $\textsf{ComposeNet}_{\lambda}$ is that the matrix composition $M_{1.j}$, which is used to derive $c_j$, depends only on $\lceil \log(j) \rceil$ rounds of matrix composition. Moreover, given that $M_{1.j}$ depends on a small subset of branches in $\textsf{ComposeNet}_{\lambda}$, it is possible to reduce the circuit significantly to extract only one bit. Call the reduced circuit that extracts the $\alpha$-th bit $\textsf{ComposeNet}^\mathsf{BTX}_{\alpha}$. This circuit can be constructed by removing nodes from $\textsf{ComposeNet}_{\lambda}$ that are not parents of $M_{1.(\alpha-1)}$. In an alternative view, it can be constructed iteratively by, at each layer, computing the pairwise matrix compositions of all nodes until the final layer contains only $M_{1.(\alpha-1)}$.

The bit extraction protocol \pbtx, described in Protocol \ref{prot:btx}, follows exactly the same structure as \pdecompopt. The only difference is that the reduced matrix composition network $\textsf{ComposeNet}^\mathsf{BTX}_{\alpha}$ is computed in lieu of $\textsf{ComposeNet}_{\lambda}$ and the depth of the circuit depends only on $\lceil \log(\alpha-1) \rceil$. An example of $\textsf{ComposeNet}^\mathsf{BTX}_{\alpha}$ for $\alpha=9$ is given in Figure \ref{fig:matcompbtx}.

\begin{procedure}[h]
    \SetKwInOut{Input}{Input}
    \SetKwInOut{Output}{Output}

    \Input{ $[\![x]\!]_{2^{\lambda}},\; \alpha  $}
    \Output{$[\![x_\alpha]\!]_2$}
    
    Party $i$ regards the lowest $\alpha$ bits of its share, $x_i$, as $p_{i,1}, \ldots,p_{i,\alpha}$ s.t. \\
    $[\![ p_j ]\!]_2, = p_{1,j} \oplus p_{2,j}$ for $j=1,\ldots, \alpha$ \\

    Party 1 creates the sharing $[\![ g_{1,j} ]\!]_2=(p_{1,j}, 0)$.\\
    Party 2 creates the sharing $[\![ g_{2,j} ]\!]_2=(0,p_{2,j})$.\\
    $[\![ g_j ]\!]_2 \leftarrow [\![ g_{1,j} ]\!]_2[\![ g_{2,j} ]\!]_2$\\
    $[\![M_j]\!]_2 \leftarrow 
        \begin{bmatrix}
            [\![p_j]\!]_2 & [\![g_j]\!]_2 \\
            0 & 1 \\
        \end{bmatrix}$ for $j=1,...,\alpha-1$\\
    $[\![M_{1.(\alpha-1)}]\!]_2 \leftarrow 
    \mathsf{ComposeNet}^\mathsf{BTX}_{\alpha}( [\![M]\!]_2 )$ \\
    $[\![c_{\alpha-1}]\!]_2 \leftarrow$ the upper right entry of $[\![M_{1.(\alpha-1)}]\!]_2$ \\

    \KwRet{$[\![p_\alpha]\!]_2 \oplus [\![c_{\alpha-1}]\!]_2$}
    \caption{Secure Protocol() \pbtx extracts the $\alpha$-th bit from a secret shared value.}\label{prot:btx}
    
\end{procedure}

\textbf{Efficiency Discussion:} Prior to computing $\textsf{ComposeNet}^\mathsf{BTX}_{\alpha}$, there is one round of communication required to compute all $g_j$. In total, this step requires $\alpha-1$ $\mathbb{Z}_2$ multiplications and a total data transfer of $2(\alpha -1)$ bits. Computing $\textsf{ComposeNet}^\mathsf{BTX}_{\alpha}$ takes $\lceil \log(\alpha-1) \rceil$ rounds of communication and, for $\alpha-1$ = $2^k$, requires 
\[ \sum\limits_{i=0}^{\log(\alpha-1)-1} 2^i = \log(\alpha-1) - 1 \] 
matrix compositions, each with 4 bits of data transfer. The complexity differs by one matrix composition for $\alpha-1 \neq 2^k$. All other operations are local and thus add virtually nothing to the overall running time. In short, the round complexity of \pbtx is $\lceil \log(\alpha-1) \rceil$ and it requires $2(\alpha - 1) + 4 \; \log(\alpha-1) - 4$ bits of data transfer.

\end{document}